\documentclass[10pt]{article}
\usepackage{graphicx}
\usepackage{amsmath}
\usepackage{amssymb}
\usepackage{caption2}
\begin{document}
\bibliographystyle{prsty}
\begin{center}
{\large {\bf \sc{  Analysis of the heavy mesons in the nuclear matter with the QCD sum rules  }}} \\[2mm]
Zhi-Gang Wang \footnote{E-mail, zgwang@aliyun.com.  }\\[2mm]
 Department of Physics, North China Electric Power University, Baoding 071003, P. R. China
\end{center}

\begin{abstract}
In this article, I calculate the contributions of the nuclear matter induced condensates up to dimension 5,  take into account the next-to-leading order contributions of the nuclear matter induced quark condensate, study the properties of the scalar, pseudoscalar, vector and axialvector heavy mesons in the nuclear matter with the QCD sum rules in a systematic way, and obtain  the shifts of the masses and decay constants. Furthermore, I study the heavy-meson-nucleon scattering lengths as a byproduct,
and obtain the conclusion qualitatively about the possible existence  of  heavy-meson-nucleon bound states.
\end{abstract}



\section{Introduction}

The suppression of  $J/\psi$ production in  relativistic  heavy ion
collisions is considered as   an important signature to identify the quark-gluon plasma
\cite{Matsui86}. The dissociation of $J/\psi$ in the quark-gluon
plasma due to color screening can result in a reduction of its
production. The interpretation of suppression requires the detailed knowledge of the
expected suppression due to the $J/\psi$ dissociation in the hadronic environment.
The  in-medium hadron properties
   can affect  the productions  of the open-charmed mesons and  the $J/\psi$ in
the relativistic heavy ion collisions,   the higher charmonium states are
considered as the major source of the $J/\psi$  \cite{Jpsi-Source}. For example,
 the higher  charmonium states can decay to the $D\bar{D}$, $D^*\bar{D}^*$ pairs instead
of decaying to the lowest  state $J/\psi$ in case of the  mass reductions
   of the $D$, $D^*$, $\bar{D}$, $\bar{D}^*$ mesons  are large enough.
 We have to  disentangle  the color screening  versus the recombination of off-diagonal $\bar{c}c$ (or $\bar{b}b$) pairs in the hot dense medium  versus  cold nuclear matter  effects, such as nuclear absorption,
shadowing and anti-shadowing, so as to draw a  definite conclusion on appearance  of the quark-gluon plasma \cite{Recombine-cc,CNM}.
The upcoming  FAIR
(Facility for Antiproton and Ion Research) project at GSI (Institute for Heavy Ion Research) in
Darmstadt (Germany)
 provides the opportunity to study the in-medium properties of the charmoniums  or charmed hadrons for the first time.
  The CBM (Compressed  Baryonic  Matter) collaboration intends to study the  properties of the
hadrons in the nuclear matter \cite{CBM}, while the $\rm \bar{P}ANDA$ (anti-Proton Annihilation at Darmstadt) collaboration will focus on the charm spectroscopy, and mass and width modifications of the charmed hadrons in the nuclear matter \cite{PANDA}.
  However,
 the in-medium mass modifications are not easy to access experimentally despite the interesting physics involved,  and they
require more detailed theoretical studies.
On the other hand, the bottomonium states  are also sensitive to the color screening, the
 $\Upsilon$ suppression in high energy heavy ion collisions can also be taken  as a
signature to identify the quark-gluon plasma \cite{QGP-rev}.
The suppressions  on the
 $\Upsilon$ production in ultra-relativistic heavy ion collisions will be studied in details  at the Relativistic Heavy Ion Collider (RHIC) and
 Large Hadron Collider (LHC).

 Extensive theoretical and experimental studies are required  to explore the  hadron properties in nuclear matter.
 The connection between the condensates and the nuclear density dependence of the in-medium hadron masses is
not straightforward.
The QCD sum rules provides  a powerful theoretical tool  in
 studying the in-medium hadronic properties \cite{SVZ79,PRT85}, and has been applied extensively
  to study the light-flavor hadrons and  charmonium states in the nuclear matter \cite{C-parameter,Drukarev1991,Jpsi-etac}.
   The
  works on the heavy mesons and heavy baryons are  few, only the $D$,  $B$, $D_0$, $B_0$, $D^*$, $B^*$, $D_1$, $B_1$,  $\Lambda_Q$,   $\Sigma_Q$, $\Xi_{QQ}$ and $\Omega_{QQ}$  are studied with the QCD sum rules \cite{Hay,WangHuang,Azi,Hil,Hilger2,WangH}.
The heavy mesons (heavy baryons) contain a  heavy quark and a light quark (two light quarks),
the existence of a light quark (two light quarks) in the heavy mesons (heavy baryons) leads to large
difference between the  mass-shifts of
 the heavy mesons (heavy baryons) and heavy quarkonia  in the nuclear matter.
The former have  large contributions  from the light-quark
condensates in the nuclear matter and the   modifications of the masses originate mainly from the modifications of the quark condensates,  while the latter
  are dominated by the gluon condensates, and the mass modifications are mild \cite{Jpsi-etac,Hay,WangHuang,Azi,Hil,Hilger2,WangH}.

  In previous works \cite{Hay,WangHuang,Azi,Hil,Hilger2},
  the properties of the heavy mesons in the nuclear matter are studied with the QCD sum rules by taking the leading order approximation for the contributions of the   quark condensates. In this article, I take into account the next-to-leading order contributions of the quark condensates,  and study the properties of the scalar, pseudoscalar, vector and axialvector  heavy mesons in the nuclear matter with the QCD sum rules in a systematic way, and make predictions for the  modifications of the masses and decay constants of the heavy mesons in the nuclear matter. Measuring those effects is a long term physics
goal based on further theoretical studies on the reaction
dynamics and on the exploration of the experimental
ability to identify more complicated
processes \cite{CBM,PANDA}.   Furthermore, I study the heavy-meson-nucleon scattering lengths as a byproduct. From the negative or positive sign of the scattering lengths, I can obtain the conclusion qualitatively that the interactions are  attractive or repulsive, which favor or disfavor  the
formations of the heavy-meson-nucleon bound states. For example, the $\Sigma_c(2800)$  and $\Lambda_c(2940)$   can  be assigned to be the $S$-wave $DN$  state with $J^P ={\frac{1}{2}}^-$ and  the $S$-wave $D^*N$  state with $J^P ={\frac{3}{2}}^-$ respectively based on the QCD sum rules \cite{ZhangDN}.

The article is arranged as follows:  I study in-medium properties of the heavy mesons
 with  the  QCD sum rules in Sec.2; in Sec.3, I present the numerical results and discussions; and Sec.4 is reserved for my
conclusions.

\section{The properties of the heavy mesons in the nuclear matter with  QCD sum rules}

I study the  scalar, pseudoscalar, vector and axialvector heavy  mesons in the nuclear matter with
the two-point correlation functions $\Pi(q)$ and $\Pi_{\mu\nu}(q)$, respectively. In the Fermi gas
approximation for the nuclear matter, I divide the $\Pi(q)$ and $\Pi_{\mu\nu}(q)$
into  the vacuum part $\Pi^0(q)$ and $\Pi^0_{\mu\nu}(q)$  and the static one-nucleon part
$\Pi_N(q)$ and $\Pi^N_{\mu\nu}(q)$,  and  expand  the
$\Pi_N(q)$ and $\Pi^N_{\mu\nu}(q)$  up to the order ${\mathcal{O}}(\rho_N)$  at relatively low nuclear density  \cite{Drukarev1991,Hay},
\begin{eqnarray}
\Pi(q) &=& i\int d^{4}x\ e^{iq \cdot x} \langle
T\left\{J(x)J^{\dag}(0)\right\} \rangle_{\rho_N}\nonumber \\
& =& \Pi_{0}(q)+ \frac{\rho_N}{2m_N}T_{N}(q)\, , \nonumber \\
 \Pi_{\mu\nu}(q) &=& i\int d^{4}x\ e^{iq \cdot x} \langle T\left\{J_\mu(x)J_\nu^{\dag}(0)\right\} \rangle_{\rho_N} \nonumber\\
 &=&\Pi^{0}_{\mu\nu}(q)+ \frac{\rho_N}{2m_N}T^{N}_{\mu\nu}(q)\, ,
 \end{eqnarray}
where the $\rho_N$ is the density of the nuclear matter, and the forward scattering amplitudes $T_{N}(q)$ and $T^{N}_{\mu\nu}(q)$ are defined as
\begin{eqnarray}
T_{N}(\omega,\mbox{\boldmath $q$}\,) &=&i\int d^{4}x e^{iq\cdot x}\langle N(p)|
T\left\{J(x)J^{\dag}(0)\right\} |N(p) \rangle\, , \nonumber\\
T^{N}_{\mu\nu}(\omega,\mbox{\boldmath $q$}\,) &=&i\int d^{4}x e^{iq\cdot x}\langle N(p)|
T\left\{J_\mu(x)J_\nu^{\dag}(0)\right\} |N(p) \rangle\, ,
\end{eqnarray}
 where the $J(x)$ and $J_\mu(x)$ denote the isospin averaged currents $\eta(x)$, $\eta_5(x)$, $\eta_\mu(x)$ and $\eta_{5\mu}(x)$, respectively,
\begin{eqnarray}
 \eta(x) &=&\eta^\dag(x) =\frac{\bar{c}(x) q(x)+\bar{q}(x) c(x)}{2}\, , \nonumber\\
  \eta_{5}(x) &=&\eta_{5}^\dag(x) =\frac{\bar{c}(x)i \gamma_5q(x)+\bar{q}(x)i\gamma_5 c(x)}{2}\,,\nonumber\\
 \eta_\mu(x) &=&\eta_\mu^\dag(x) =\frac{\bar{c}(x)\gamma_\mu q(x)+\bar{q}(x)\gamma_\mu c(x)}{2}\, , \nonumber\\
  \eta_{5\mu}(x) &=&\eta_{5\mu}^\dag(x) =\frac{\bar{c}(x)\gamma_\mu \gamma_5q(x)+\bar{q}(x)\gamma_\mu\gamma_5 c(x)}{2}\,,
\end{eqnarray}
 which interpolate the scalar, pseudoscalar, vector and axialvector mesons $D_0$, $D$, $D^*$ and $D_1$, respectively.
 I choose the isospin averaged currents since the $D_0$, $D$, $D^*$ and $D_1$ mesons are produced in pairs in the antiproton-nucleon
annihilation processes.
The $q$ denotes the $u$ or $d$ quark, the $q^{\mu}=(\omega,\mbox{\boldmath $q$}\,)$ is the four-momentum carried by
the  currents $J(x)$ and $J_\mu(x)$, the $|N(p)\rangle$ denotes the isospin and
spin averaged static nucleon state with the four-momentum $p =
(m_N,0)$, and  $\langle N(\mbox{\boldmath $p$})|N(\mbox{\boldmath
$p$}')\rangle = (2\pi)^{3} 2p_{0}\delta^{3}(\mbox{\boldmath
$p$}-\mbox{\boldmath $p$}')$ \cite{Hay}.

I can decompose the correlation functions $T^{N}_{\mu\nu}(\omega,\mbox{\boldmath $q$}\,)$ as
\begin{eqnarray}
T^{N}_{\mu\nu}(\omega,\mbox{\boldmath $q$}\,) &=&T_{N}(\omega,\mbox{\boldmath $q$}\,)\left(-g_{\mu\nu}+\frac{q_\mu q_\nu}{q^2}\right)+T^0_N(\omega,\mbox{\boldmath $q$}\,) q_\mu q_\nu\nonumber\\
&&+T^1_N(\omega,\mbox{\boldmath $q$}\,) \left(q_\mu u_\nu+q_\nu u_\mu \right)+ T^2_N(\omega,\mbox{\boldmath $q$}\,) u_\mu u_\nu\, ,
\end{eqnarray}
according to Lorentz covariance, where the $T_{N}(\omega,\mbox{\boldmath $q$}\,)$ denotes the contributions of the vector and axialvector charmed mesons,
 and the $T^{0/1/2}_{N}(\omega,\mbox{\boldmath $q$}\,)$ are irrelevant in the present analysis.

In the limit $\mbox{\boldmath $q$}\rightarrow {\bf 0}$, the forward scattering amplitude
$T_{N}(\omega,\mbox{\boldmath $q$}\,)$ can be related to the $DN$ ($D_0N$, $D^*N$ and $D_1N$)
scattering $T$-matrix,
\begin{eqnarray}
{{\cal T}_{D/D_0/D^*/D_1\,N}}(m_{D/D_0/D^*/D_1},0) =
8\pi(m_N+m_{D/D_0/D^*/D_1})a_{D/D_0/D^*/D_1} \, ,
\end{eqnarray}
 where the  $a_{D/D_0/D^*/D_1}$ are the $D/D_0/D^*/D_1\,N$
scattering lengths.  I can parameterize  the
phenomenological spectral densities $\rho(\omega,0)$ with three unknown   parameters $a,\,b$ and $c$ near the pole positions of the charmed mesons $D$, $D_0$, $D^*$ and $D_1$ according to Ref.\cite{Hay},
\begin{eqnarray}
\rho(\omega,0) &=& -\frac{1}{\pi} \mbox{Im} \left[\frac{{{\cal T}_{D/D_0N}}(\omega,{\bf 0})}{\left(\omega^{2}-
m_{D/D_0}^2+i\varepsilon\right)^{2}} \right]\frac{f_{D/D_0}^2m_{D/D_0}^4}{m_c^2}+ \cdots \,, \nonumber \\
&=& a\,\frac{d}{d\omega^2}\delta\left(\omega^{2}-m_{D/D_0}^2\right) +
b\,\delta\left(\omega^{2}-m_{D/D_0}^2\right) + c\,\delta\left(\omega^{2}-s_{0}\right)\, ,
\end{eqnarray}
for the pseudoscalar and scalar currents $\eta_5(x)$ and $\eta(x)$,
\begin{eqnarray}
\rho(\omega,0) &=& -\frac{1}{\pi} \mbox{Im} \left[\frac{{{\cal T}_{D^*/D_1N}}(\omega,{\bf 0})}{\left(\omega^{2}-
m_{D^*/D_1}^2+i\varepsilon\right)^{2}} \right]f_{D^*/D_1}^2m_{D^*/D_1}^2+ \cdots \,, \nonumber\\
&=& a\,\frac{d}{d\omega^2}\delta\left(\omega^{2}-m_{D^*/D_1}^2\right) +
b\,\delta\left(\omega^{2}-m_{D^*/D_1}^2\right) + c\,\delta\left(\omega^{2}-s_{0}\right)\, ,
\end{eqnarray}
for the vector and axialvector currents $\eta_\mu(x)$ and $\eta_{5\mu}(x)$.

Now the hadronic correlation functions $\Pi(\omega,0)$  and $\Pi_{\mu\nu}(\omega,0)$ at the phenomenological side can be written as
\begin{eqnarray}
\Pi(\omega,0)&=&\frac{\left( f_{D/D_0}+\delta f_{D/D_0}\right)^2\left( m_{D/D_0}+\delta m_{D/D_0}\right)^4}{m_c^2}\frac{1}{\left(m_{D/D_0}+\delta m_{D/D_0}\right)^2-\omega^2}+\cdots  \nonumber\\
&=&\frac{ f_{D/D_0}^2m_{D/D_0}^4}{m_c^2}\frac{1}{m_{D/D_0}^2-\omega^2}+\cdots\nonumber\\
&&+\frac{\rho_N}{2m_N}\left[\frac{a}{\left(m_{D/D_0}^2-\omega^2\right)^2}+\frac{b}{m_{D/D_0}^2-\omega^2} +\cdots \right]\, ,
\end{eqnarray}
\begin{eqnarray}
\Pi_{\mu\nu}(\omega,0)&=&\left( f_{D^*/D_1}+\delta f_{D^*/D_1}\right)^2\left( m_{D^*/D_1}+\delta m_{D^*/D_1}\right)^2\frac{1}{\left(m_{D^*/D_1}+\delta m_{D^*/D_1}\right)^2-\omega^2}\nonumber\\
&&\left( -g_{\mu\nu}+\frac{q_\mu q_\nu}{q^2}\right) +\cdots \, ,\nonumber\\
&=&f_{D^*/D_1}^2m_{D^*/D_1}^2\frac{1}{m_{D^*/D_1}^2-\omega^2}\left( -g_{\mu\nu}+\frac{q_\mu q_\nu}{q^2}\right)+\cdots\nonumber\\
&&+\frac{\rho_N}{2m_N}\left[\left(\frac{a}{\left(m_{D^*/D_1}^2-\omega^2\right)^2}+\frac{b}{m_{D^*/D_1}^2-\omega^2} +\cdots\right)\left( -g_{\mu\nu}+\frac{q_\mu q_\nu}{q^2}\right)+\cdots \right] \, .\nonumber\\
\end{eqnarray}

In Eqs.(6-7), the first
term denotes the double-pole term, and corresponds  to the on-shell
effect of the $T$-matrix,
\begin{eqnarray}
a&=&-8\pi(m_N+m_{D/D_0})a_{D/D_0}\frac{f_{D/D_0}^2m_{D/D_0}^4}{m_c^2}\, ,
\end{eqnarray}
for the currents $\eta_5(x)$ and $\eta(x)$ and
\begin{eqnarray}
a&=&-8\pi(m_N+m_{D^*/D_1})a_{D^*/D_1}f_{D^*/D_1}^2m_{D^*/D_1}^2\, ,
\end{eqnarray}
for the currents  $\eta_\mu(x)$ and $\eta_{5\mu}(x)$;
the second term denotes  the single-pole term,
and corresponds to the off-shell effect of the $T$-matrix;
 the
third term denotes   the continuum term or the  remaining effects,
where the $s_{0}$ is the continuum threshold parameter.
In general, the continuum contributions are approximated by $\rho_{QCD}(\omega,0)\theta(\omega^2-s_0)$, where the $\rho_{QCD}(\omega,0)$ are the
perturbative QCD spectral densities, and $\theta(x)=1$ for $x\geq 0$, else $\theta(x)=0$. In this article, the QCD spectral densities are of the type $\delta(\omega^2-m_Q^2)$, which include both the ground state and continuum state contributions,  I have attributed  the excited state contributions to the continuum state contributions,  so the collective continuum state contributions can be approximated as $c\,\delta(\omega^2-s_0)$, then I obtain
 the result $c/\left(s_0-\omega^2\right)$ in the hadronic representation, see Eq.(15). The doublet $\left(D(2550), D(2600)\right)$ or $\left(D_J(2580), D^*_J (2650)\right)$ is assigned  to be
 the first radial excited state of the doublet $(D,D^*)$ \cite{WangHQET}. The single-pole contributions come from the doublet $\left(D(2550), D(2600)\right)$ or $\left(D_J(2580), D^*_J (2650)\right)$ are  of the form $1/\left(m_{D(2550)/D(2600)}^2-\omega^2\right)$, so the approximation $c/\left(s_0-\omega^2\right)$ is reasonable.

Then the shifts of the masses and decay  constants  of
the charmed-mesons can be approximated as
\begin{eqnarray}
\delta m_{D/D_0/D^*/D_1} &=&2\pi\frac{m_{N}+m_{D/D_0/D^*/D_1}}{m_Nm_{D/D_0/D^*/D_1}}\rho_N a_{D/D_0/D^*/D_1}\, ,
\end{eqnarray}
\begin{eqnarray}
\delta f_{D/D_0}&=&\frac{m_c^2}{2f_{D/D_0}m_{D/D_0}^4}\left(\frac{b\rho_N}{2m_N}-\frac{4f_{D/D_0}^2m_{D/D_0}^3\delta m_{D/D_0}}{m_c^2} \right) \, , \nonumber\\
\delta f_{D^*/D_1}&=&\frac{1}{2f_{D^*/D_1}m_{D^*/D_1}^2}\left(\frac{b\rho_N}{2m_N}-2f_{D^*/D_1}^2m_{D^*/D_1}\delta m_{D^*/D_1} \right) \, .
\end{eqnarray}

In calculations, I have used the following definitions for the decay constants of the heavy mesons,
\begin{eqnarray}
\langle 0|\eta(0)|D_0+\bar{D}_0\rangle &=&\frac{f_{D_0}m^2_{D_0}}{m_c} \,,\nonumber\\
\langle 0|\eta_5(0)|D+\bar{D}\rangle &=&\frac{f_{D}m^2_{D}}{m_c} \,,\nonumber\\
\langle 0|\eta_\mu(0)|D^*+\bar{D}^*\rangle &=&f_{D^*}m_{D^*}\epsilon_\mu\,,\nonumber\\
\langle 0|\eta_{5\mu}(0)|D_1+\bar{D}_1\rangle &=&f_{D_1}m_{D_1}\epsilon_{\mu}\,,
\end{eqnarray}
with summations of the polarization vectors  $\sum_\lambda \epsilon_\mu(\lambda,q)\epsilon^*_\nu(\lambda,q)=-g_{\mu\nu}+\frac{q_\mu q_\nu}{q^2}$.

In the low energy limit $\omega\rightarrow 0$, the
$T_{N}(\omega,{\bf 0})$ is equivalent to the Born term $T_{N}^{\rm
Born}(\omega,{\bf 0})$. Now I take
into account the Born terms  at the phenomenological side,
\begin{eqnarray}
T_{N}(\omega^2)&=&T_{N}^{\rm
Born}(\omega^2)+\frac{a}{\left(m_{D/D_0/D^*/D_1}^2-\omega^2\right)^2}+\frac{b}{m_{D/D_0/D^*/D_1}^2-\omega^2}+\frac{c}{s_0-\omega^2}
\, ,
\end{eqnarray}
with the constraint
\begin{eqnarray}
\frac{a}{m_{D/D_0/D^*/D_1}^4}+\frac{b}{m_{D/D_0/D^*/D_1}^2}+\frac{c}{s_0}&=&0 \, .
\end{eqnarray}

The contributions from the intermediate spin-$\frac{3}{2}$ charmed
baryon states are zero in the soft-limit $q_\mu \to 0$
\cite{Wangzg}, and I only take into account the intermediate
spin-$\frac{1}{2}$ charmed  baryon states in calculating the Born terms,
\begin{eqnarray}
\left(D/D_0/D^*/D_1\right)^0(c\bar{u})+p(uud)\ \mbox{or}\ n(udd)&\longrightarrow&
\Lambda_c^+,\Sigma_c^+(cud)\ \mbox{or}\ \Sigma_c^0(cdd) \, ,\nonumber\\
 \left(D/D_0/D^*/D_1\right)^+(c\bar{d})+p(uud)\ \mbox{or}\ n(udd)
&\longrightarrow& \Sigma_c^{++}(cuu)\ \mbox{or}\
\Lambda_c^+,\Sigma_c^+(cud)\, ,
\end{eqnarray}
where   $M_{\Lambda_c}=2.286\,\rm{GeV}$   and
$M_{\Sigma_c}=2.454\,\rm{GeV}$ \cite{PDG}. I can take $M_H\approx
2.4\,\rm{GeV}$ as the average value, where the $H$ means either
$\Lambda_c^+$, $\Sigma_c^+$, $\Sigma_c^{++}$ or $\Sigma_c^0$. In the case of the bottom baryons, I take the approximation $M_H=\frac{M_{\Sigma_b}+M_{\Lambda_b}}{2}\approx 5.7\,\rm{GeV}$ \cite{PDG}. I write down the Feynman  diagrams and calculate the  Born terms directly,
and obtain    the  results,
\begin{eqnarray}
T_{N}^{\rm
Born}(\omega,{\bf0})&=&\frac{2m_N(m_H+m_N)}
{\left[\omega^2-(m_H+m_N)^2\right]\left[\omega^2-m_{D}^2\right]^2}\left(\frac{f_{D}m_{D}^2g_{DNH}}{m_c}\right)^2\,,
\end{eqnarray}
for the current $\eta_5(x)$,
\begin{eqnarray}
T_{N}^{\rm
Born}(\omega,{\bf0})&\to&T_{N}^{\rm
Born}(\omega,{\bf0})\,\left( {\rm with}\,\,\, m_N \to -m_N\, , \,\,\,  D\to D_0 \right)\,,
\end{eqnarray}
for the current $\eta(x)$,

\begin{eqnarray}
T_{N}^{\rm
Born}(\omega,{\bf0})&=&\frac{2m_N(m_H+m_N)}
{\left[\omega^2-(m_H+m_N)^2\right]\left[\omega^2-m_{D^*}^2\right]^2}\left(f_{D^*}m_{D^*}g_{D^*NH}\right)^2\,,
\end{eqnarray}
for the current $\eta_\mu(x)$,
\begin{eqnarray}
T_{N}^{\rm
Born}(\omega,{\bf0})&\to&T_{N}^{\rm
Born}(\omega,{\bf0})\,\left( {\rm with}\,\,\, m_N \to -m_N\, , \,\,\,  D^*\to D_1 \right)\,,
\end{eqnarray}
for the current $\eta_{5\mu}(x)$, where the $g_{D/D_0/D^*/D_1NH}$ denote the  strong coupling constants
$g_{D/D_0/D^*/D_1N\Lambda_c}$
 and  $g_{D/D_0/D^*/D_1N\Sigma_c}$.
On the other hand, there are no inelastic channels for the
$(\bar{D}/\bar{D}_0/\bar{D}^*/\bar{D}_1)^0  N$ and $(\bar{D}/\bar{D}_0/\bar{D}^*/\bar{D}_1)^- N$ interactions, and $T_{N}^{\rm
Born}(0)=0$.
In calculations, I have used the following definitions for  the hadronic coupling constants,
 \begin{eqnarray}
 \langle\Lambda_c/\Sigma_c(p-q)|D(-q)N(p)\rangle &=&g_{\Lambda_c/\Sigma_cDN}\overline{U}_{\Lambda_c/\Sigma_c}(p-q) i\gamma_5 U_N(p)\,,\nonumber\\
 \langle\Lambda_c/\Sigma_c(p-q)|D_0(-q)N(p)\rangle &=& g_{\Lambda_c/\Sigma_cD_0N}\overline{U}_{\Lambda_c/\Sigma_c}(p-q) U_N(p)\,,\nonumber\\
 \langle\Lambda_c/\Sigma_c(p-q)|D^*(-q)N(p)\rangle &=&\overline{U}_{\Lambda_c/\Sigma_c}(p-q)\left( g_{\Lambda_c/\Sigma_cD^*N}\!\not\!{\epsilon}+i\frac{g^T_{\Lambda_c/\Sigma_cD^*N}}{M_N+M_{\Lambda_c/\Sigma_c}}\sigma^{\alpha\beta}\epsilon_\alpha q_\beta\right)U_N(p)\,,\nonumber\\
 \langle\Lambda_c/\Sigma_c(p-q)|D_1(-q)N(p)\rangle &=&\overline{U}_{\Lambda_c/\Sigma_c}(p-q)\left( g_{\Lambda_c/\Sigma_cD_1N}\!\not\!{\epsilon}+i\frac{g^T_{\Lambda_c/\Sigma_cD_1N}}{M_N+M_{\Lambda_c/\Sigma_c}}\sigma^{\alpha\beta}\epsilon_\alpha q_\beta\right)\gamma_5U_N(p)\,,\nonumber\\
 \end{eqnarray}
 where the $U_N$ and $\overline{U}_{\Lambda_c/\Sigma_c}$ are the Dirac spinors of the nucleon and the charmed  baryons $\Lambda_c/\Sigma_c$, respectively. In the limit $q_\mu \to 0$, the  strong coupling constants $g_{\Lambda_c/\Sigma_cD^*N}^T$ and  $g_{\Lambda_c/\Sigma_cD_1N}^T$ have no contributions.

 For example, near the thresholds, the $D^*N$  can translate  to the $DN$, $D^*N$, $\pi\Sigma_c$, $\eta\Lambda_c$,   etc, we can  take into account  the intermediate baryon-meson loops or the re-scattering effects  with the Bethe-Salpeter equation to obtain the full $D^*N \to D^*N$ scattering amplitude, and   generate higher baryon states  dynamically \cite{DN-BSE-Negative}. We can saturate the full $D^*N\to D^*N$ scattering amplitude with the tree-level Feynman diagrams describing   the  exchanges of the higher resonances $\Lambda_c(2595)$, $\Sigma_c({\frac{1}{2}}^-)$, etc.  While in other coupled-channels analysis,  the $\Lambda_c(2595)$ emerges as a $DN$ quasi-bound state rather than a $D^*N$ quasi-bound state \cite{DN-BSE-Negative}.     The translations  $D^*N $ to the ground  states $\Lambda_c $ and $\Sigma_c$ are favored in the phase-space, as   the $\Lambda_c(2595)$ and $\Sigma_c({\frac{1}{2}}^-)$ with $J^P={\frac{1}{2}}^-$  have the average mass $m_{H'}\approx 2.7\,\rm{GeV}$  \cite{PDG,Wang-Negative}. In fact,  $m_{H'}^2> s_0$, I can absorb the high resonances into the continuum states in case the high resonances do not dominate the
 QCD sum rules.
 In calculations, I observe that the mass-shift $\delta m_{D^*}$ does not sensitive to contributions of the  ground  states $\Lambda_c $ and $\Sigma_c$,
  the contributions from the spin-$\frac{1}{2}$ higher resonances maybe even smaller.
In this article, I neglect the intermediate baryon-meson loops, their  effects are absorbed into continuum contributions.

At the low nuclear density,  the condensates $\langle{\cal {O}}\rangle_{\rho_N}$ in the nuclear matter can be approximated as
\begin{eqnarray}
\langle{\cal{O}}\rangle_{\rho_N} &=&\langle{\cal{O}}\rangle+\frac{\rho_N}{2m_N}\langle {\cal{O}}\rangle_N \, ,
\end{eqnarray}
based on the Fermi gas model, where the   $\langle{\cal{O}}\rangle$ and $\langle
{\cal{O}}\rangle_N$ denote the vacuum condensates and nuclear matter induced condensates,  respectively \cite{Drukarev1991}. I neglect the terms
proportional to $p_F^4$, $p_F^5$, $p_F^6$, $\cdots$
at the normal nuclear matter with the saturation density
$\rho_N=\rho_0=\frac{2p_F^3}{3\pi^2}$, as the Fermi momentum
$p_F=0.27\,\rm{GeV}$ is a small quantity \cite{Drukarev1991}.

I carry out the operator product expansion to the  nuclear matter induced condensates $\frac{\rho_N}{2m_N}\langle
{\cal{O}}\rangle_N$ up to dimension-5
 at the large space-like  region in the nuclear matter,
and take into account the one-loop corrections to the quark condensate $\langle \bar{q}q\rangle_N$. I insert the following term
\begin{eqnarray}
\frac{1}{2!}\,\, ig_s \int d^D y \bar{\psi}(y)\gamma^\mu \psi(y)\frac{\lambda^a}{2}G^a_\mu(y)\,\, ig_s \int d^D z \bar{\psi}(z)\gamma^\nu \psi(z)\frac{\lambda^b}{2}G^b_\nu(z) \, ,
\end{eqnarray}
 with the dimension  $D=4-2\epsilon$, into the correlation functions $T_N(q)$ and $T^N_{\mu\nu}(q)$ firstly, where the $\psi$ denotes the quark fields, the $G^a_\mu$  denotes the gluon field, the $\lambda^a$ denotes  the Gell-Mann matrix, then contract the quark fields with Wick theorem,  and extract the quark condensate $\langle\bar{q}{q}\rangle_N$ according to the formula  $\langle N|q_\alpha^i q_\beta^j|N\rangle=-\frac{1}{12}\langle \bar{q}q\rangle_N\delta_{ij}\delta_{\alpha\beta}$ to obtain the perturbative corrections $\alpha_s\langle\bar{q}{q}\rangle_N$, where the $i$ and $j$ are color indexes and the $\alpha$ and $\beta$ are Dirac spinor  indexes. There are six Feynman diagrams make contributions, see Fig.1. Now I calculate  the first diagram explicitly for the current $\eta_5(x)$ in Fig.1,
 \begin{eqnarray}
 2T^{(\alpha_s,1)}_N(q^2) &=&-\frac{{\rm Tr}(\frac{\lambda^a}{2} \frac{\lambda^b}{2})\langle \bar{q}q\rangle_Ng_s^2 \mu^{2\epsilon}}{12} \frac{i}{(2\pi)^D}\int d^Dk{\rm Tr}\left\{ i\gamma_5 \frac{i}{\!\not\! {k}}\gamma^\alpha \gamma^\beta \frac{i}{\!\not\! {k}} i \gamma_5 \frac{i}{\!\not\! {q}+\!\not\! {k}-m_c} \frac{-i\delta_{ab}g_{\alpha\beta}}{k^2}\right\} \nonumber\\
 &=&-\frac{4Dm_c\langle \bar{q}q\rangle_Ng_s^2 \mu^{2\epsilon}}{3(2\pi)^D}i\frac{\partial}{\partial t}\frac{(-2\pi i)^2}{2\pi i} \int_{m_c^2}^{\infty} ds \frac{\int d^Dk \delta \left( k^2-t\right)\delta\left( (k+q)^2-m_c^2\right)}{s-q^2}\mid_{t=0} \nonumber\\
  &=& -\frac{Dm_c\langle \bar{q}q\rangle_Ng_s^2 \mu^{2\epsilon}\left[1+\epsilon(\log4\pi-\gamma_E) \right]}{12\pi^2}  \int_{m_c^2}^{\infty} ds \frac{1}{s-q^2} \frac{s+m_c^2}{s^{1-\epsilon}(s-m_c^2)^{1+2\epsilon}} \, ,
   \end{eqnarray}
   where I have used Cutkosky's rule to obtain the QCD spectral density. There exists  infrared divergence at the end point $s=m_c^2$.
   It is difficult to carry out the integral over $s$, I can perform the Borel transform $B_{M^2}$ firstly, then carry out the  integral over $s$,
 \begin{eqnarray}
 B_{M^2}2T^{(\alpha_s,1)}_N(q^2)&=& -\frac{Dm_c\langle \bar{q}q\rangle_Ng_s^2 \mu^{2\epsilon}\left[1+\epsilon(\log4\pi-\gamma_E) \right]}{12\pi^2M^2}  \int_{m_c^2}^{\infty} ds   \frac{s+m_c^2}{s^{1-\epsilon}(s-m_c^2)^{1+2\epsilon}}\exp\left( -\frac{s}{M^2}\right) \nonumber\\
 &=&\frac{m_c\langle \bar{q}q\rangle_Ng_s^2}{3\pi^2M^2}\exp\left(-\frac{m_c^2}{M^2} \right)\left( \frac{1}{\epsilon}-\log4\pi+\gamma_E\right)+\frac{m_c\langle \bar{q}q\rangle_Ng_s^2}{3\pi^2M^2}\Gamma\left(0,\frac{m_c^2}{M^2} \right) \nonumber\\
 &&-\frac{m_c\langle \bar{q}q\rangle_Ng_s^2}{6\pi^2M^2}\exp\left(-\frac{m_c^2}{M^2} \right)+\frac{m_c\langle \bar{q}q\rangle_Ng_s^2}{3\pi^2M^2}\exp\left(-\frac{m_c^2}{M^2} \right)\log\frac{m_c^2\mu^2}{M^4} \, ,
 \end{eqnarray}
 where
 \begin{eqnarray}
\Gamma(0,x)&=&e^{-x}\int_0^\infty dt \frac{1}{t+x}e^{-t} \, .
\end{eqnarray}
 Other diagrams are calculated analogously,   I regularize the divergences in $D=4-2\epsilon$ dimension, then remove the ultraviolet divergences through renormalization and absorb the
    infrared divergences into the quark condensate $\langle \bar{q}q\rangle_N$.
 \begin{figure}
 \centering
 \includegraphics[totalheight=6cm,width=14cm]{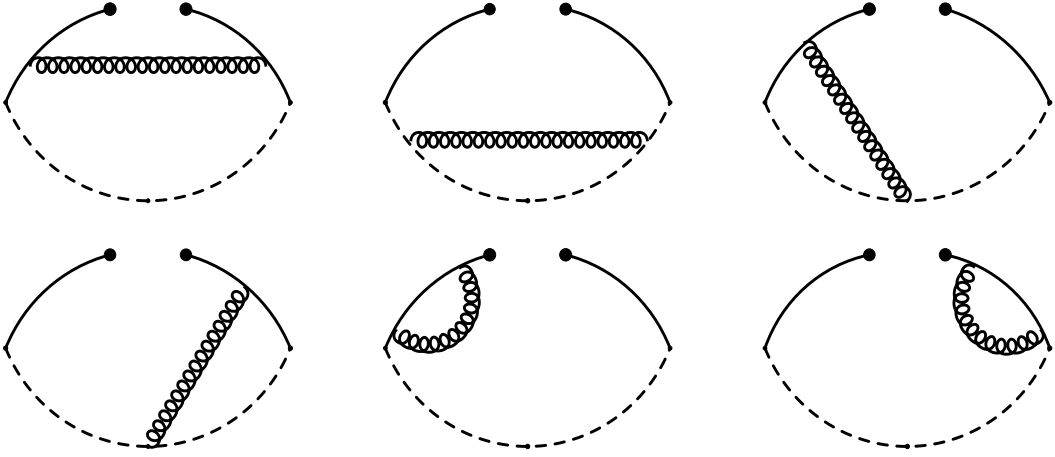}
    \caption{The perturbative $\mathcal{O}(\alpha_s)$ corrections to the quark condensate $\langle\bar{q}q\rangle_N$. }
\end{figure}

I calculate the contributions of other condensates at the tree level, the calculations are straightforward and  cumbersome. In calculations, I use the following formulas,
\begin{eqnarray}
\langle
q_{\alpha}(x)\bar{q}_{\beta}(0)\rangle_N&=&-\frac{1}{4}\left[\left(\langle\bar{q}q\rangle_N+x^{\mu}\langle\bar{q}D_{\mu}q\rangle_N
+\frac{1}{2}x^{\mu}x^{\nu}\langle\bar{q}D_{\mu}D_{\nu}q\rangle_N
+\cdots\right)\delta_{\alpha\beta}\right.\nonumber \\
&&\left.+\left(\langle\bar{q}\gamma_{\lambda}q\rangle_N+x^{\mu}\langle\bar{q}
\gamma_{\lambda}D_{\mu} q\rangle_N
+\frac{1}{2}x^{\mu}x^{\nu}\langle\bar{q}\gamma_{\lambda}D_{\mu}D_{\nu}
q\rangle_N
+\cdots\right)\gamma^{\lambda}_{\alpha\beta} \right] \, ,
\end{eqnarray}
 and
\begin{eqnarray}
\langle
g_{s}q^i_{\alpha}\bar{q}^j_{\beta}G_{\mu\nu}^{a}\rangle_N&=&-\frac{1}{96} \frac{\lambda^a_{ij}}{2}\left\{\langle g_{s}\bar{q}\sigma Gq\rangle_N\left[\sigma_{\mu\nu}+i(u_{\mu}\gamma_{\nu}-u_{\nu}\gamma_{\mu
})
\!\not\! {u}\right]_{\alpha\beta} +\langle g_{s}\bar{q}\!\not\! {u}\sigma Gq\rangle_N\right.\nonumber\\
&&\left.\left[\sigma_{\mu\nu}\!\not\!
{u}+i(u_{\mu}\gamma_{\nu}-u_{\nu}\gamma_{\mu}
)\right]_{\alpha\beta}
-4\langle\bar{q}u D u D q\rangle_N\left[\sigma_{\mu\nu}+2i(u_{\mu}\gamma_{\nu}-u_{\nu}\gamma_{\mu}
)\!\not\! {u}\right]_{\alpha\beta}\right\} \, , \nonumber\\
\end{eqnarray}
where  $D_\mu=\partial_\mu-ig_s\frac{\lambda^a}{2}G^a_\mu$,
\begin{eqnarray} \label{ }
\langle\bar{q}\gamma_{\mu}q\rangle_N&=&\langle\bar{q}\!\not\!{u}q\rangle_N
u_{\mu} \, , \nonumber \\
\langle\bar{q}D_{\mu}q\rangle_N&=&\langle\bar{q}u D
q\rangle_N
u_{\mu}=0\, , \nonumber \\
\langle\bar{q}\gamma_{\mu}D_{\nu}q\rangle_N&=&\frac{4}{3}\langle\bar{q}
\!\not\! {u}u D q\rangle_N\left(u_{\mu}u_{\nu}-\frac{1}{4}g_{\mu\nu}\right) \, , \nonumber \\
\langle\bar{q}D_{\mu}D_{\nu}q\rangle_N&=&\frac{4}{3}\langle\bar{q}
u D u D q\rangle_N\left(u_{\mu}u_{\nu}-\frac{1}{4}g_{\mu\nu}\right)
-\frac{1}{6} \langle
g_{s}\bar{q}\sigma Gq\rangle_N\left(u_{\mu}u_{\nu}-g_{\mu\nu}\right) \, , \nonumber \\
\langle\bar{q}\gamma_{\lambda}D_{\mu}D_{\nu}q\rangle_N&=&2\langle\bar{q}
\!\not\! {u}u  D u  D
q\rangle_N\left[u_{\lambda}u_{\mu}u_{\nu} -\frac{1}{6}
\left(u_{\lambda}g_{\mu\nu}+u_{\mu}g_{\lambda\nu}+u_{\nu}g_{\lambda\mu}\right)\right]
\nonumber\\
&&-\frac{1}{6} \langle
g_{s}\bar{q}\!\not\! {u}\sigma Gq\rangle_N(u_{\lambda}u_{\mu}u_{\nu}-u_{\lambda}g_{\mu\nu}) \, ,
\end{eqnarray}
and
\begin{equation}
 \langle
G_{\alpha\beta}^{a}G_{\mu\nu}^{b}\rangle_N=\frac{\delta^{ab}}{96}
\langle
GG\rangle_N\left(g_{\alpha\mu}g_{\beta\nu}-g_{\alpha\nu}g_{\beta\mu}\right)+O\left(\langle
\textbf{E}^{2}+\textbf{B}^{2}\rangle_N\right).
\end{equation}

Once analytical results at the level of quark-gluon degree's of  freedom are obtained,
 then I set  $\omega^2=q^2$, and take the
quark-hadron duality below the continuum threshold $s_0$, and perform the Borel transform  with respect
to the variable $Q^2=-\omega^2$, finally   obtain  the following QCD sum
rules:
\begin{eqnarray}
 a\, C_a+b\, C_b &=&C_f \, ,
\end{eqnarray}
\begin{eqnarray}
C_a &=&\frac{1}{M^2}\exp\left(-\frac{m_{D}^2}{M^2}\right)-\frac{s_0}{m_{D}^4}\exp\left(-\frac{s_0}{M^2}\right) \, ,\nonumber\\
C_b&=&\exp\left(-\frac{m_{D}^2}{M^2}\right)-\frac{s_0}{m_{D}^2}\exp\left(-\frac{s_0}{M^2}\right) \, ,
\end{eqnarray}

\begin{eqnarray}
C_f&=& \frac{2m_N(m_H+m_N)}{(m_H+m_N)^2-m_{D}^2}\left(\frac{f_{D}m_{D}^2g_{DNH}}{m_c}\right)^2\left\{ \left[\frac{1}{M^2}-\frac{1}{m_{D}^2-(m_H+m_N)^2}\right] \exp\left(-\frac{m_{D}^2}{M^2}\right)\right.\nonumber\\
&&\left.+\frac{1}{(m_H+m_N)^2-m_{D}^2}\exp\left(-\frac{(m_H+m_N)^2}{M^2}\right)\right\}-\frac{m_c\langle\bar{q}q\rangle_N}{2}\left\{1+\frac{\alpha_s}{\pi} \left[ 6-\frac{4m_c^2}{3M^2} \right.\right.\nonumber\\
&&\left.\left.-\frac{2}{3}\left( 1-\frac{m_c^2}{M^2}\right)\log\frac{m_c^2}{\mu^2}-2\Gamma\left(0,\frac{m_c^2}{M^2}\right)\exp\left( \frac{m_c^2}{M^2}\right) \right]\right\}\exp\left(- \frac{m_c^2}{M^2}\right) \nonumber\\
&&+\frac{1}{2}\left\{-2\left(1-\frac{m_c^2}{M^2}\right)\langle q^\dag i D_0q\rangle_N +\frac{4m_c
}{M^2}\left(1-\frac{m_c^2}{2M^2}\right)\langle \bar{q} i D_0 i D_0q\rangle_N+\frac{1}{12}\langle\frac{\alpha_sGG}{\pi}\rangle_N\right\} \nonumber\\
&&\exp\left(- \frac{m_c^2}{M^2}\right)\, ,
\end{eqnarray}
 for the current $\eta_5(x)$,
 \begin{eqnarray}
 C_i &\to& C_i\left( {\rm with}\,\,\, m_N \to -m_N\, , \,\,\,m_c \to -m_c\, , \,\,\,D \to D_0\right)\, ,
 \end{eqnarray}
for the current $\eta(x)$,
\begin{eqnarray}
C_a &=&\frac{1}{M^2}\exp\left(-\frac{m_{D^*}^2}{M^2}\right)-\frac{s_0}{m_{D^*}^4}\exp\left(-\frac{s_0}{M^2}\right) \, ,\nonumber\\
C_b&=&\exp\left(-\frac{m_{D^*}^2}{M^2}\right)-\frac{s_0}{m_{D^*}^2}\exp\left(-\frac{s_0}{M^2}\right) \, ,
\end{eqnarray}
\begin{eqnarray}
C_f&=& \frac{2m_N(m_H+m_N)}{(m_H+m_N)^2-m_{D^*}^2}\left(f_{D^*}m_{D^*}g_{D^*NH}\right)^2\left\{ \left[\frac{1}{M^2}-\frac{1}{m_{D^*}^2-(m_H+m_N)^2}\right] \exp\left(-\frac{m_{D^*}^2}{M^2}\right)\right.\nonumber\\
&&\left.+\frac{1}{(m_H+m_N)^2-m_{D^*}^2}\exp\left(-\frac{(m_H+m_N)^2}{M^2}\right)\right\}-\frac{m_c\langle\bar{q}q\rangle_N}{2}\left\{1+\frac{\alpha_s}{\pi} \left[ \frac{8}{3}-\frac{4m_c^2}{3M^2} \right.\right.\nonumber\\
&&\left.\left.+\frac{2}{3}\left( 2+\frac{m_c^2}{M^2}\right)\log\frac{m_c^2}{\mu^2}-\frac{2m_c^2}{3M^2}\Gamma\left(0,\frac{m_c^2}{M^2}\right)\exp\left( \frac{m_c^2}{M^2}\right) \right]\right\}\exp\left(- \frac{m_c^2}{M^2}\right) \nonumber\\
&&+\frac{1}{2}\left\{-\frac{4\langle q^\dag i D_0q\rangle_N}{3} +\frac{2m_c^2\langle q^\dag i D_0q\rangle_N}{M^2}+\frac{2m_c\langle\bar{q}g_s\sigma Gq\rangle_N}{3M^2}+\frac{16m_c\langle \bar{q} i D_0 i D_0q\rangle_N}{3M^2}\right.\nonumber\\
&&\left.-\frac{2m_c^3\langle \bar{q} i D_0 i D_0q\rangle_N}{M^4}-\frac{1}{12}\langle\frac{\alpha_sGG}{\pi}\rangle_N\right\}\exp\left(- \frac{m_c^2}{M^2}\right)\, ,
\end{eqnarray}
 for the current $\eta_\mu(x)$,
\begin{eqnarray}
 C_i &\to& C_i\left( {\rm with}\,\,\, m_N \to -m_N\, , \,\,\,m_c \to -m_c\, , \,\,\,D^* \to D_1\right)\, ,
 \end{eqnarray}
for the current $\eta_{5\mu}(x)$, where $i=a,b,f$.
In this article, I neglect the contributions from the heavy quark condensates $\langle \bar{Q}Q\rangle$, $\langle \bar{Q}Q\rangle=-\frac{1}{12\pi m_Q}\langle \frac{\alpha_s GG}{\pi} \rangle$ up to the order $\mathcal{O}(\alpha_s)$ (here I count the condensate $ \langle \frac{\alpha_s GG}{\pi} \rangle$ as of the order $\mathcal{O}(\alpha_s)$), the heavy quark condensates have practically no effect on the polarization
functions, for detailed discussions about this subject, one can consult Ref.\cite{PRT85}.
In Ref.\cite{QQcond}, Buchheim, Hilger and Kampfer study the contributions of the  condensates involve the heavy quarks in details, the results indicate that
those condensates are either suppressed by the heavy quark mass $m_Q$ or by the additional factor $\frac{\alpha_s}{4\pi}$ (or $g_s^2/(4\pi)^2$).
Neglecting  the in-medium effects on the heavy quark condensates cannot affect the predictions remarkably, as the main contributions come from the terms $\langle \bar{q}q\rangle_N$.

Differentiate  above equation with respect to  $\tau=\frac{1}{M^2}$, then
eliminate the
 parameter $b$ ($a$), I can obtain the QCD sum rules for
 the parameter $a$ ($b$),
 \begin{eqnarray}
 a&=&\frac{C_f\left(-\frac{d}{d\tau}\right)C_b-C_b\left(-\frac{d}{d\tau}\right)C_f}{C_a\left(-\frac{d}{d\tau}\right)C_b-C_b\left(-\frac{d}{d\tau}\right)C_a}\, , \nonumber\\
  b&=&\frac{C_f\left(-\frac{d}{d\tau}\right)C_a-C_a\left(-\frac{d}{d\tau}\right)C_f}{C_b\left(-\frac{d}{d\tau}\right)C_a-C_a\left(-\frac{d}{d\tau}\right)C_b}\, .
 \end{eqnarray}
 With the simple replacements $m_c \to m_b$, $D/D_0/D^*/D_1 \to B/B_0/B^*/B_1$, $\Lambda_c \to \Lambda_b$ and $\Sigma_c \to \Sigma_b$,
 I can obtain the corresponding the QCD sum rules for  the bottom  mesons in the nuclear matter.

\section{Numerical results and discussions}
At the normal nuclear matter with the saturation density
$\rho_N=\rho_0=\frac{2p_F^3}{3\pi^2}$, where  the Fermi momentum
$p_F=0.27\,\rm{GeV}$ is a small quantity,
  the condensates $\langle{\cal {O}}\rangle_{\rho_N}$ in the nuclear matter can be approximated as
$\langle{\cal{O}}\rangle_{\rho_N} =\langle{\cal{O}}\rangle+\frac{\rho_N}{2m_N}\langle {\cal{O}}\rangle_N $,   the terms
proportional to $p_F^4$, $p_F^5$, $p_F^6$, $\cdots$ can be neglected safely,
where the   $\langle{\cal{O}}\rangle=\langle0|{\cal{O}}|0\rangle$ and $\langle
{\cal{O}}\rangle_N=\langle N|
{\cal{O}}|N\rangle$ denote the vacuum condensates and nuclear matter induced condensates,  respectively \cite{Drukarev1991}.

The  input parameters at the QCD side are taken as $\rho_N=(0.11\,\rm{GeV})^3$,
  $\langle\bar{q} q\rangle_N={\sigma_N \over m_u+m_d } (2m_N)$,
 $\langle\frac{\alpha_sGG}{\pi}\rangle_N= - 0.65 \,{\rm {GeV}} (2m_N)$, $\sigma_N=45\,\rm{MeV}$,
 $m_u+m_d=12\,\rm{MeV}$,
$\langle q^\dagger iD_0 q\rangle_N=0.18 \,{\rm{GeV}}(2m_N)$,
$\langle\bar{q}g_s\sigma G q\rangle_N=3.0\,{\rm GeV}^2(2m_N) $,
$\langle \bar{q} iD_0iD_0
q\rangle_N+{1\over8}\langle\bar{q}g_s\sigma G
q\rangle_N=0.3\,{\rm{GeV}}^2(2m_N)$,
 $m_N=0.94\,\rm{GeV}$ \cite{C-parameter}, $m_c=(1.3\pm0.1)\,\rm{GeV}$, $m_b=(4.7\pm0.1)\,\rm{GeV}$, $\alpha_s=0.45$ and $\mu=1\,\rm{GeV}$.
  If we take the normalization $\langle N(\mbox{\boldmath $p$})|N(\mbox{\boldmath
$p$}')\rangle = (2\pi)^{3} \delta^{3}(\mbox{\boldmath
$p$}-\mbox{\boldmath $p$}')$, then $\langle{\cal{O}}\rangle_{\rho_N} =\langle{\cal{O}}\rangle+\rho_N\langle {\cal{O}}\rangle_N $, the unit $2m_N$ in the brackets in the values of  the condensates $\langle\bar{q} q\rangle_N$, $\langle\frac{\alpha_sGG}{\pi}\rangle_N$, $\cdots$ disappears.
I choose the values of the nuclear matter  induced condensates determined in Ref.\cite{C-parameter}, which are still widely used in the literatures.
Although the values of some condensates are updated, those condensates are irrelevant to the present work. The updates focus  on
the four-quark condensate \cite{update-4q}. In this article, I take into account the condensates up to dimension-5, the four-quark condensates have no contributions, the dominant  contributions come from the nuclear matter induced condensate $\langle\bar{q} q\rangle_N$,  $\langle\bar{q} q\rangle_N={\sigma_N \over m_u+m_d } (2m_N)$. The value $m_u+m_d=12\,\rm{MeV}$ is obtained from the famous Gell-Mann-Oakes-Renner relation at the energy scale $\mu=1\,\rm{GeV}$, while the value $\sigma_N=45\,\rm{MeV}$ is still widely used \cite{update-4q}.

The parameters at the hadronic side are taken as
$m_D=1.870\,\rm{GeV}$, $m_B=5.280\,\rm{GeV}$, $m_{D_0}=2.355\,\rm{GeV}$, $m_{B_0}=5.740\,\rm{}GeV$,
 $m_{D^*}=2.010\,\rm{GeV}$, $m_{B^*}=5.325\,\rm{GeV}$,  $m_{D_1}=2.420\,\rm{GeV}$, $m_{B_1}=5.750\,\rm{}GeV$,
$f_{D}=0.210\,\rm{GeV}$, $f_B=0.190\,\rm{GeV}$,
 $f_{D_0}=0.334 \frac{m_c}{m_{D_0}}\,\rm{GeV}$,
$f_{B_0}=0.280\frac{m_b}{m_{B_0}}\,\rm{GeV}$,
$f_{D^*}=0.270\,\rm{GeV}$, $f_{B^*}=0.195\,\rm{GeV}$,
 $f_{D_1}=0.305\,\rm{GeV}$,
$f_{B_1}=0.255\,\rm{GeV}$,
 $s^0_{D}=(6.2\pm0.5)\,\rm{GeV}^2$, $s^0_{B}=(33.5\pm1.0)\,\rm{GeV}^2$,
$s^0_{D^*}=(6.5\pm0.5)\,\rm{GeV}^2$, $s^0_{B^*}=(35.0\pm1.0)\,\rm{GeV}^2$,
$s^0_{D_0}=(8.0\pm0.5)\,\rm{GeV}^2$, $s^0_{B_0}=(39.0\pm1.0)\,\rm{GeV}^2$,
$s^0_{D_1}=(8.5\pm0.5)\,\rm{GeV}^2$ and $s^0_{B_1}=(39.0\pm1.0)\,\rm{GeV}^2$, which are determined by the conventional two-point correlation functions
 using the QCD sum rules \cite{WangHuang,WangJHEP}. I neglect the uncertainties of the decay constants to avoid double  counting as the main uncertainties of the decay constants originate from the uncertainties of the continuum threshold parameters $s_0$.

The value of the strong coupling constant $g_{DN\Lambda_c}$ is $g_{\Lambda_c DN}=6.74$ from
the  QCD sum rules \cite{Nielsen98}, while the average value of the strong coupling constants $g_{\Lambda_cDN}$ and $g_{\Sigma_cDN}$ from the light-cone QCD sum rules is $\frac{g_{\Lambda_cDN}+g_{\Sigma_cDN}}{2}=6.775$ \cite{Khodjamirian1108}, those values are consistent with each other.
The average value of the strong coupling constants $g_{\Lambda_c D^*N}$ and $g_{\Sigma_c D^*N}$ from the light-cone QCD sum rules is  $\frac{g_{\Lambda_c D^*N}+g_{\Sigma_cD^*N}}{2}=3.86$ \cite{Khodjamirian1108}.  In this
article, I take the approximation
$g_{DN\Lambda_c}\approx
g_{DN\Sigma_c}\approx g_{BN\Lambda_b}\approx
g_{BN\Sigma_b}\approx g_{D_0N\Lambda_c}\approx
g_{D_0N\Sigma_c}\approx g_{B_0N\Lambda_b}\approx
g_{B_0N\Sigma_b}\approx6.74$ and
$g_{\Lambda_cD^*N}\approx g_{\Sigma_cD^*N}
\approx g_{\Lambda_cD_1N}\approx g_{\Sigma_cD_1N}\approx g_{\Lambda_bB^*N}\approx g_{\Sigma_bB^*N}
\approx g_{\Lambda_bB_1N}\approx g_{\Sigma_bB_1N}\approx3.86$.

\begin{figure}
 \centering
 \includegraphics[totalheight=5cm,width=7cm]{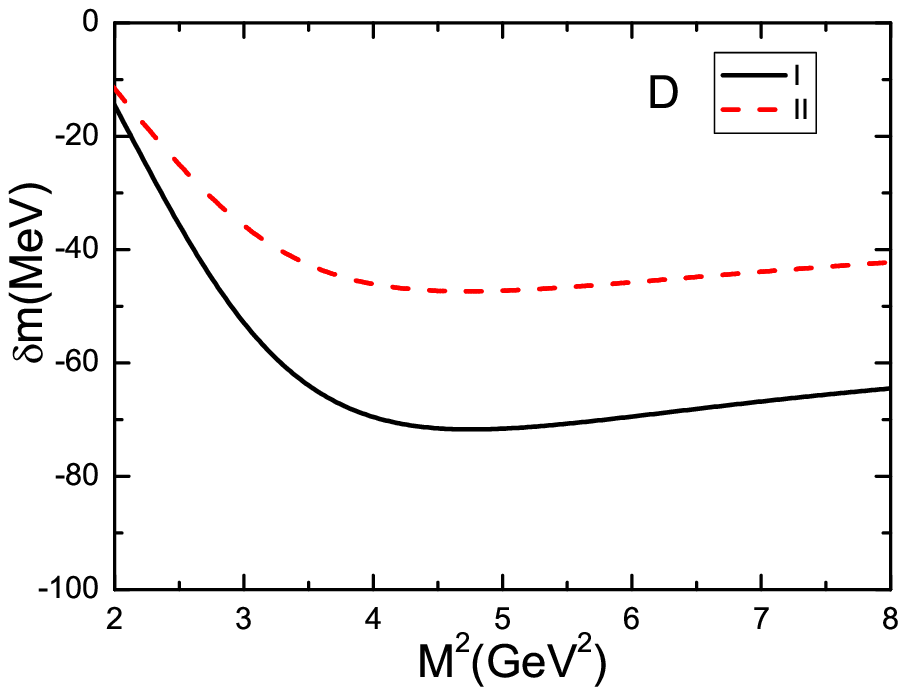}
  \includegraphics[totalheight=5cm,width=7cm]{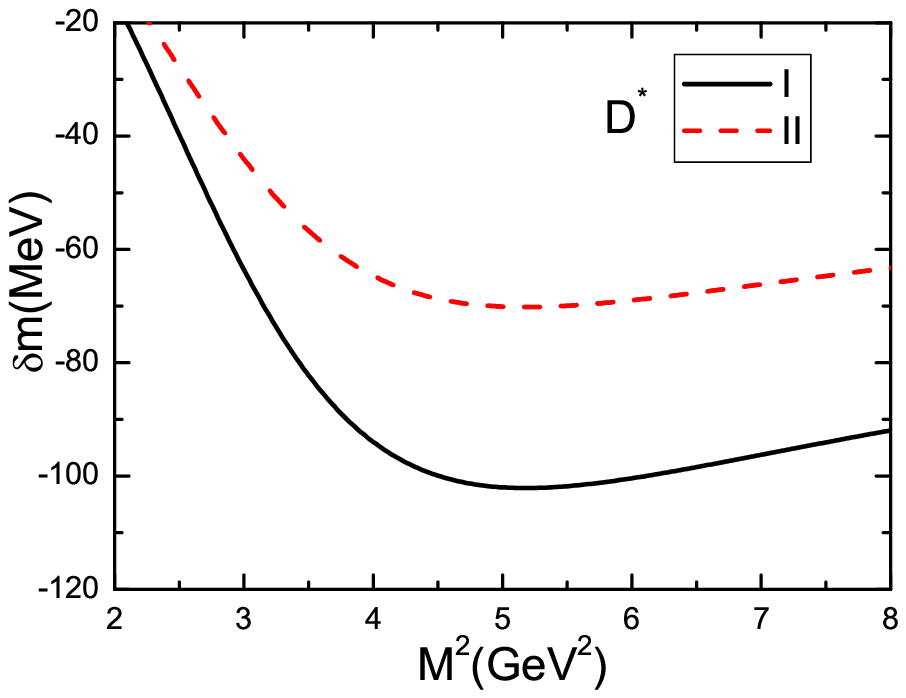}
  \includegraphics[totalheight=5cm,width=7cm]{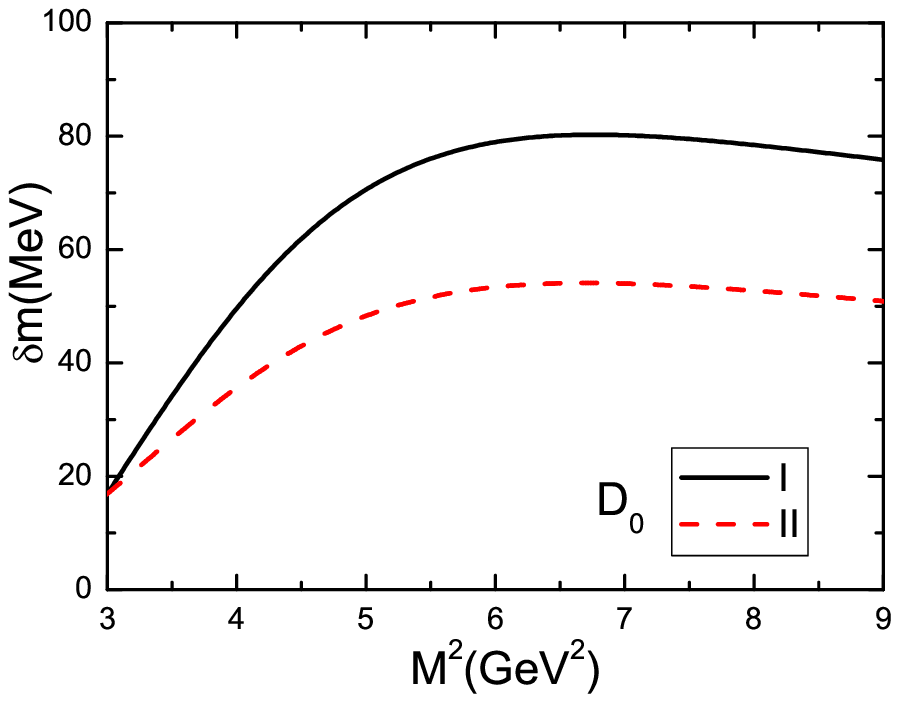}
  \includegraphics[totalheight=5cm,width=7cm]{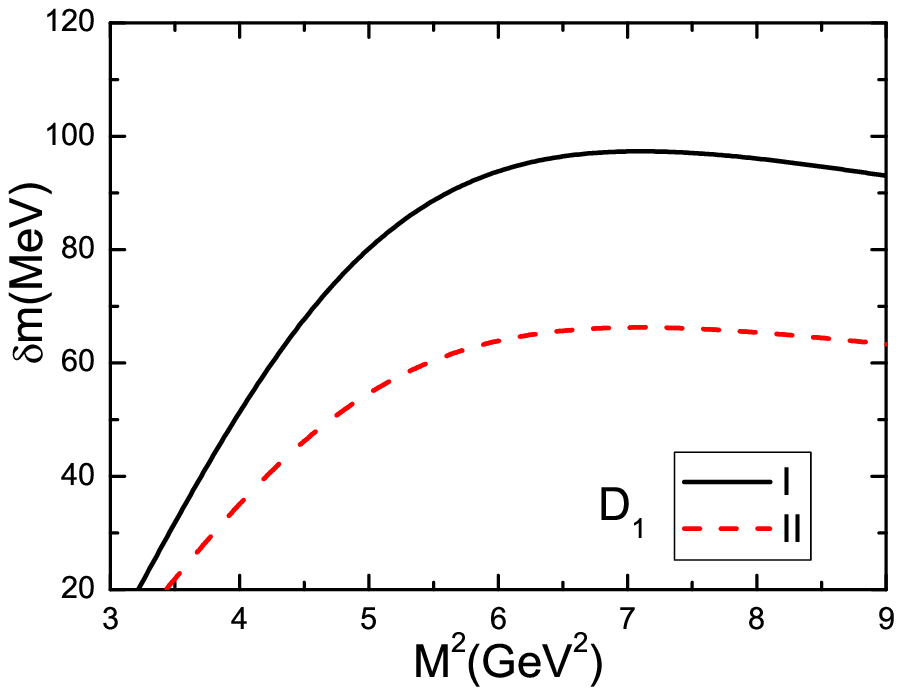}
   \includegraphics[totalheight=5cm,width=7cm]{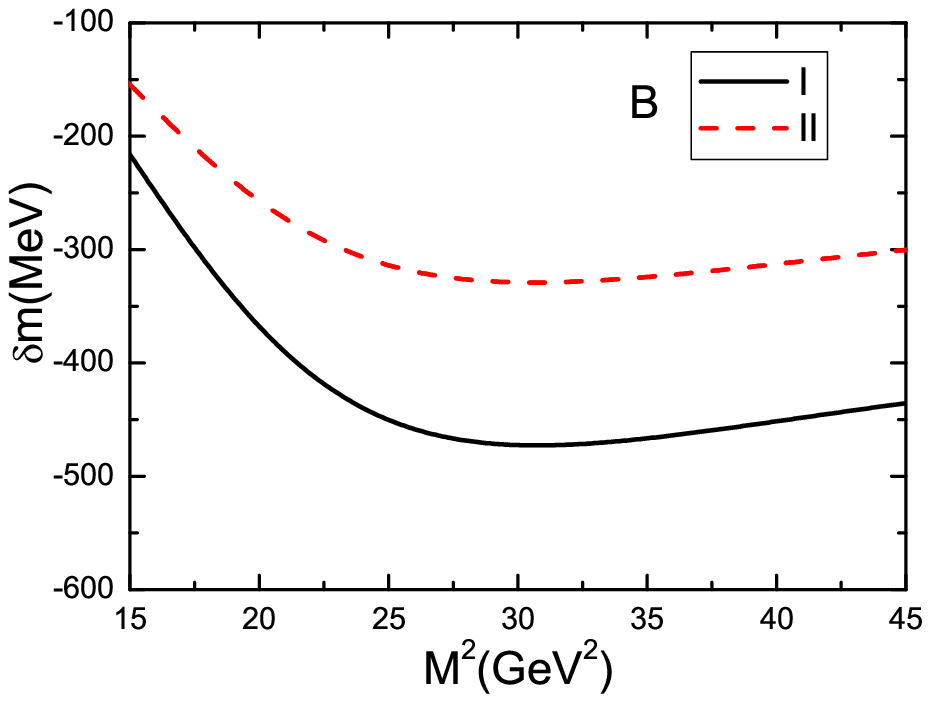}
    \includegraphics[totalheight=5cm,width=7cm]{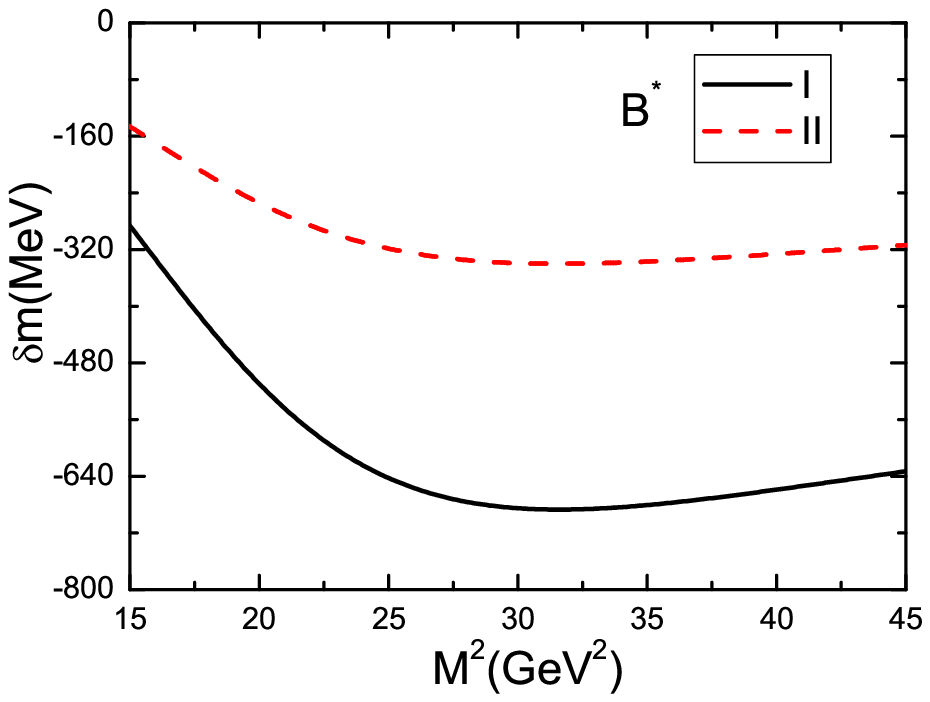}
    \includegraphics[totalheight=5cm,width=7cm]{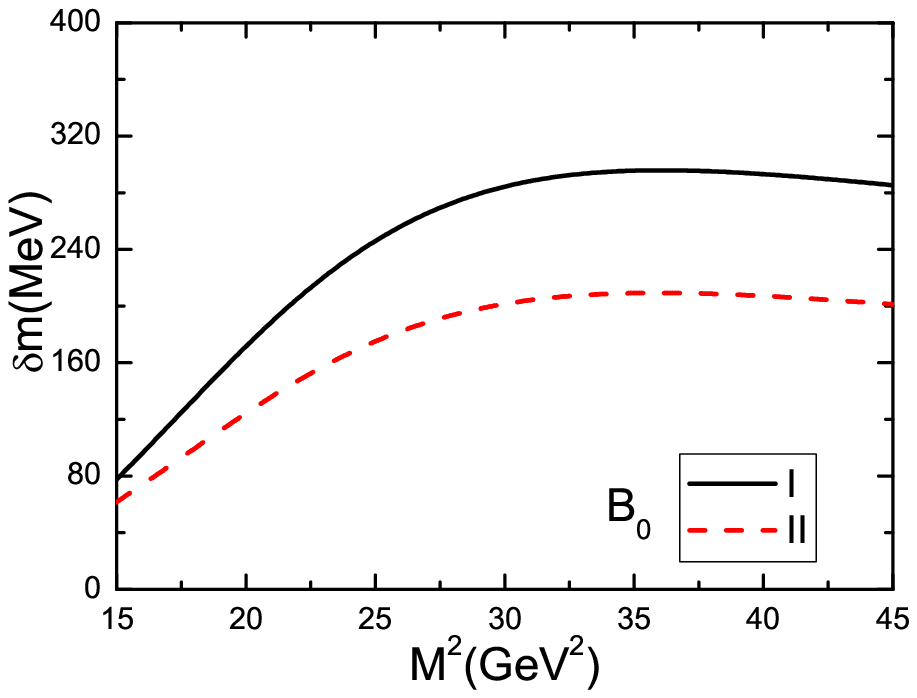}
    \includegraphics[totalheight=5cm,width=7cm]{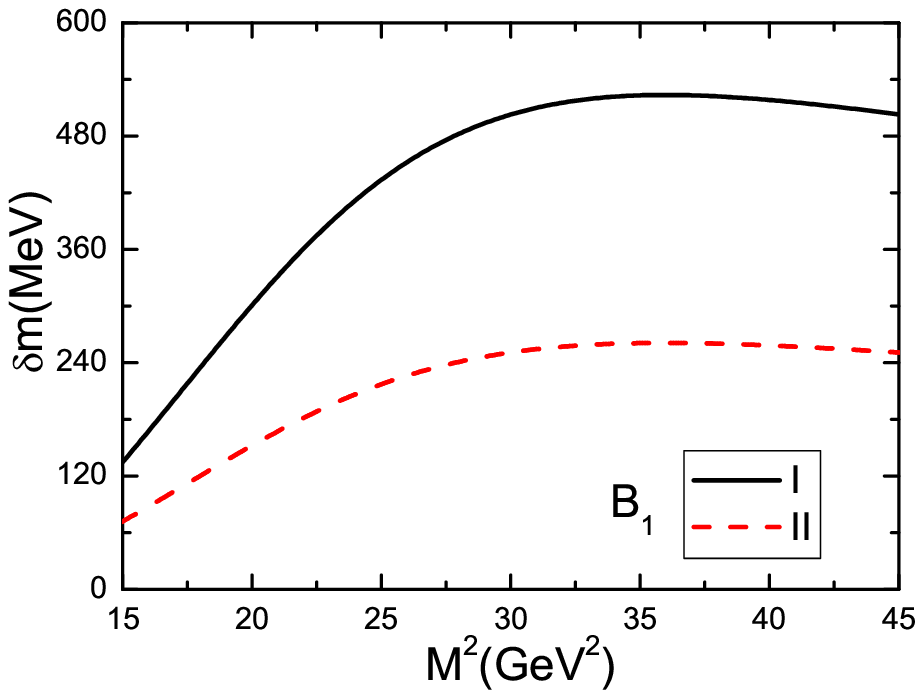}
 \caption{(Color online) The shifts of the masses  of the heavy mesons in the nuclear matter with variations  of the Borel parameter $M^2$, the I (II) denotes contributions up to  the next-to-leading order (leading order) are included. }
\end{figure}

\begin{figure}
 \centering
 \includegraphics[totalheight=5cm,width=7cm]{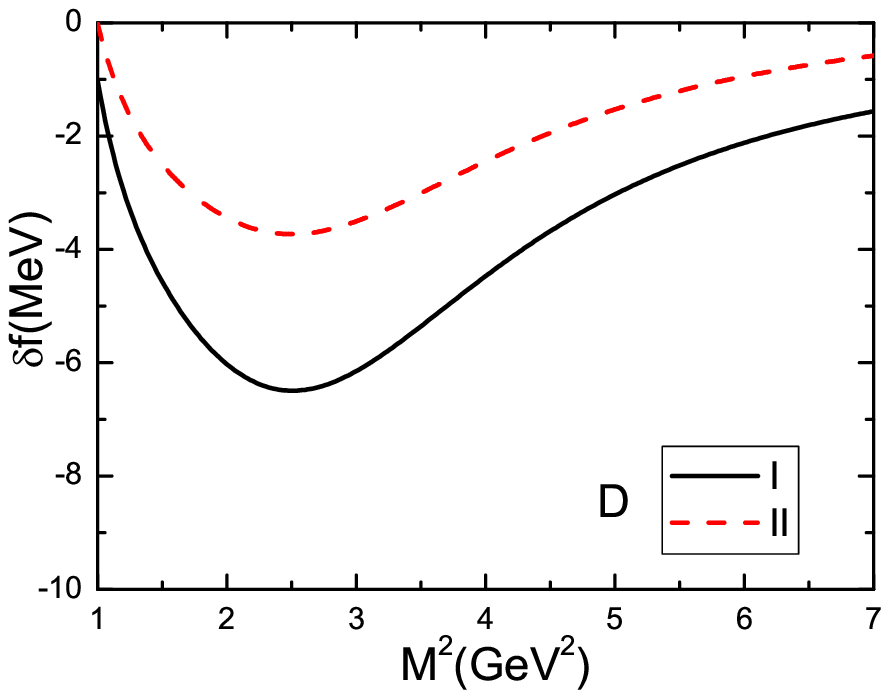}
  \includegraphics[totalheight=5cm,width=7cm]{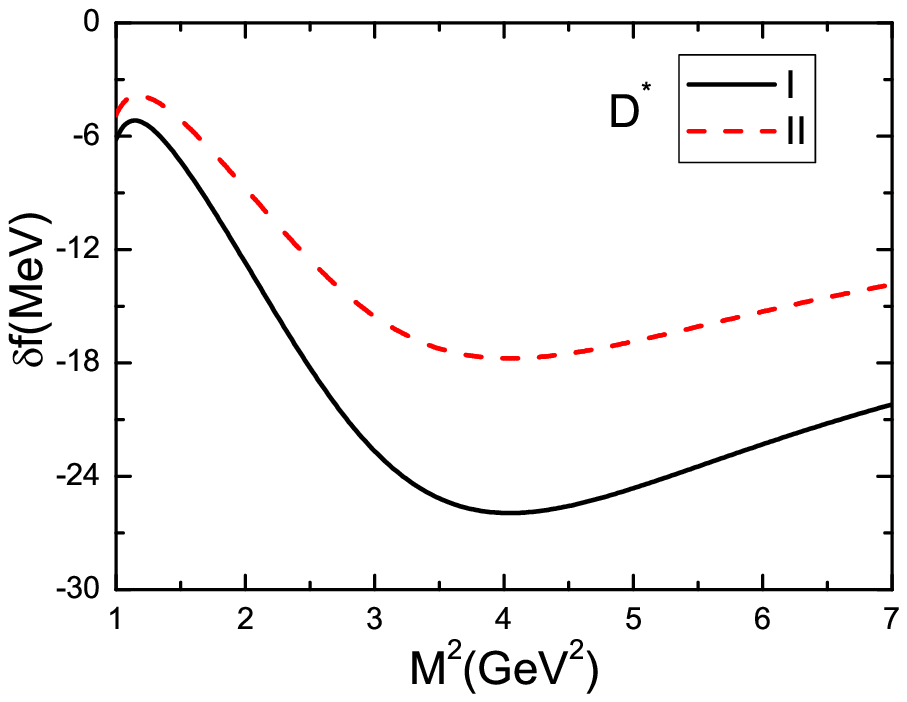}
  \includegraphics[totalheight=5cm,width=7cm]{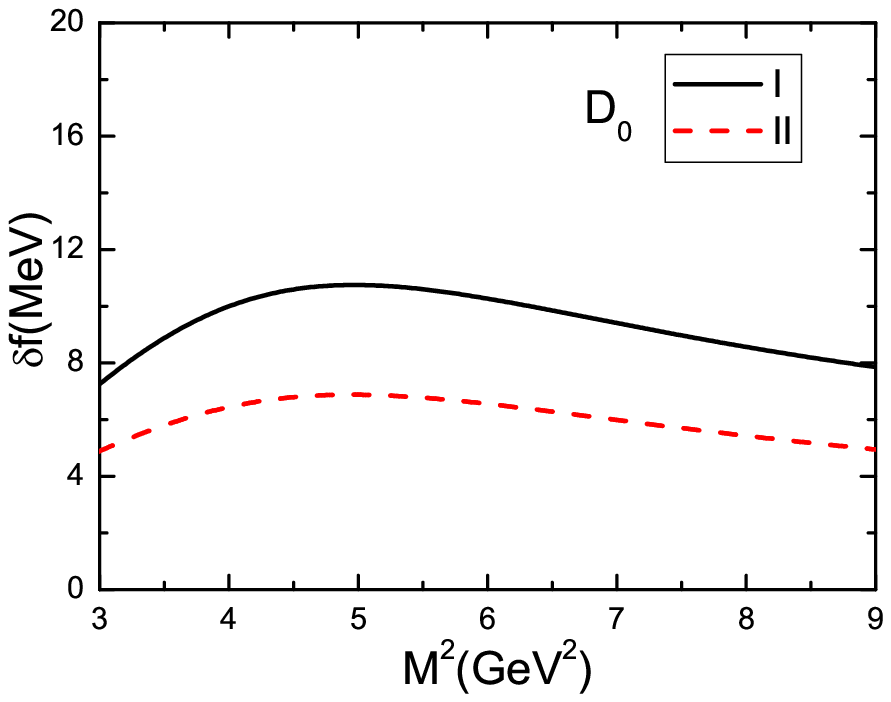}
  \includegraphics[totalheight=5cm,width=7cm]{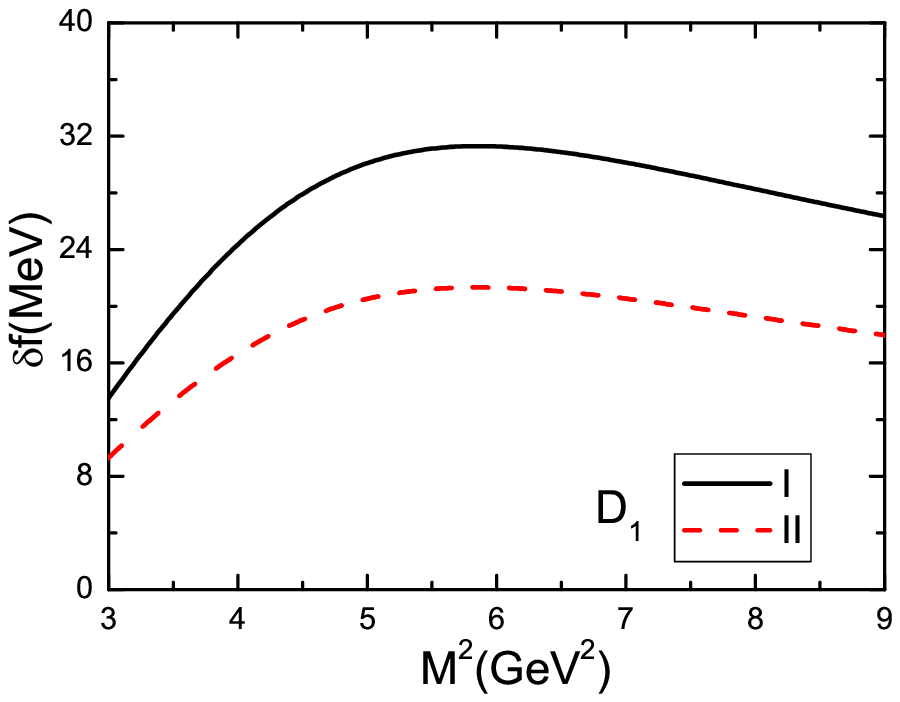}
  \includegraphics[totalheight=5cm,width=7cm]{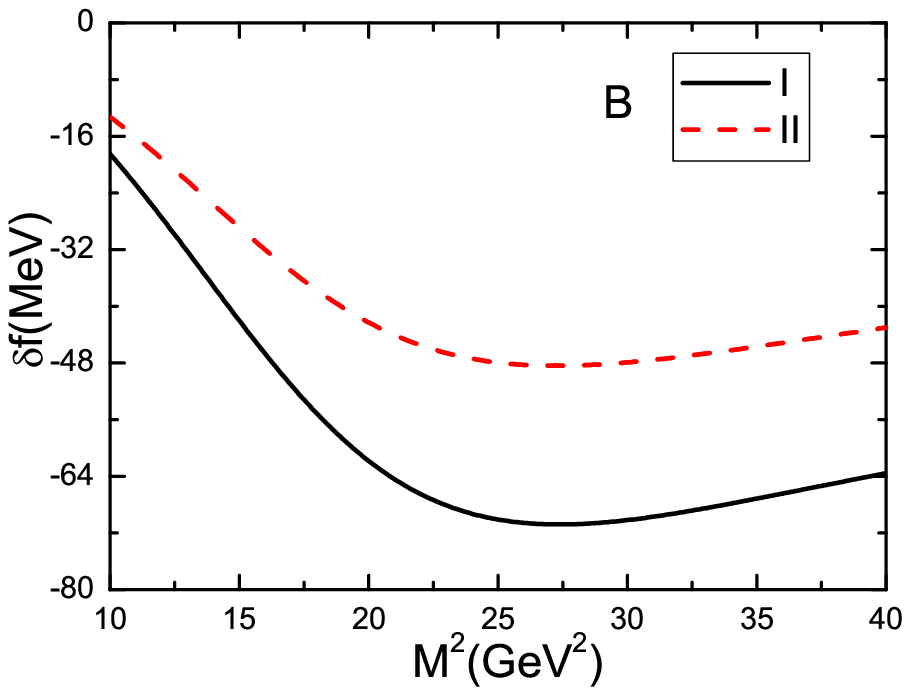}
  \includegraphics[totalheight=5cm,width=7cm]{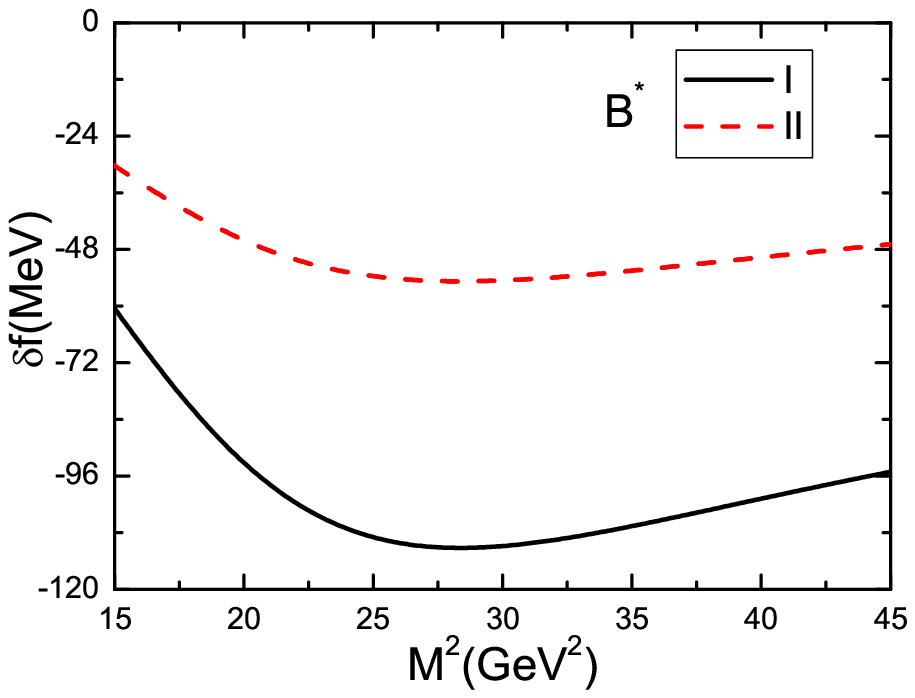}
   \includegraphics[totalheight=5cm,width=7cm]{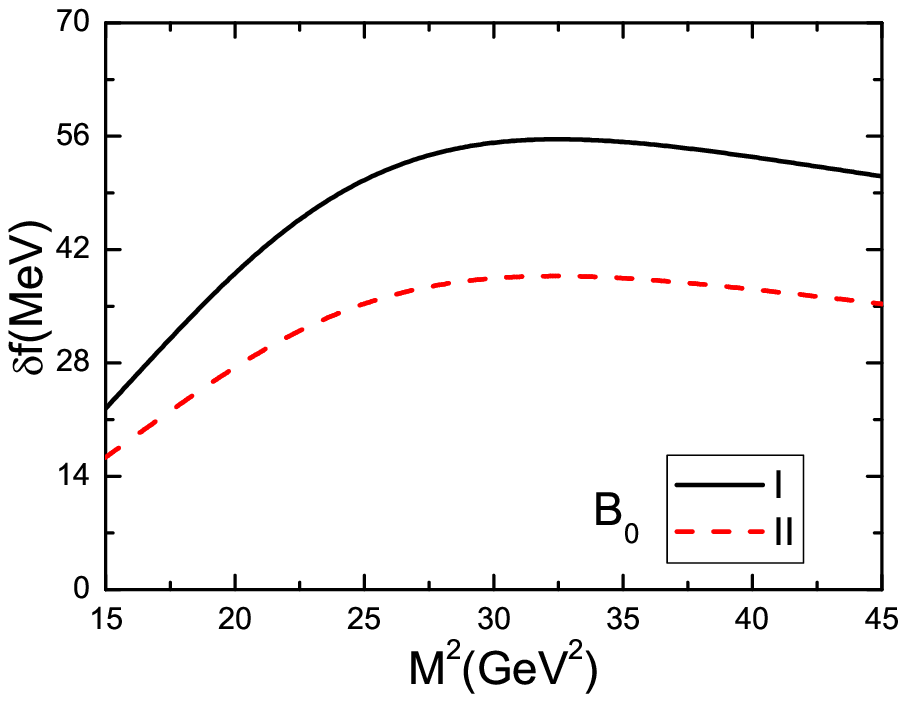}
    \includegraphics[totalheight=5cm,width=7cm]{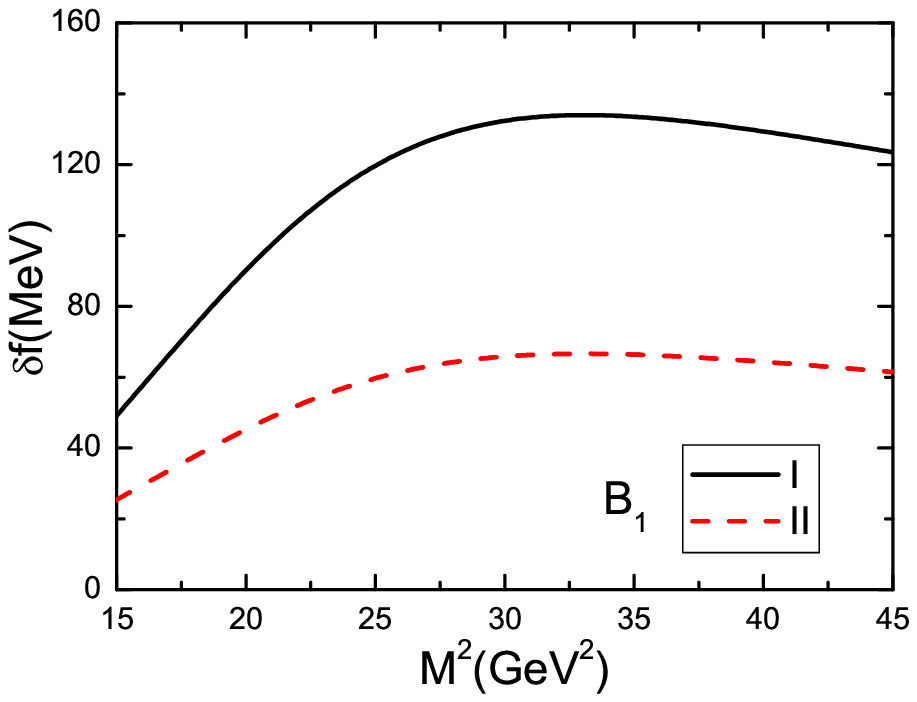}
 \caption{(Color online) The shifts of the decay constants  of the heavy mesons in the nuclear matter with variations  of the Borel parameter $M^2$, the I (II) denotes contributions up to  the next-to-leading order (leading order) are included. }
\end{figure}

In Figs.2-3, I plot the shifts of the masses and decay constants of the heavy mesons   in the nuclear matter with variations  of  the Borel
parameter $M^2$, respectively. From the figures, I can see that there appear platforms. In this article, I choose the Borel parameters $M^2$ according to the criterion that the uncertainties originate from the Borel parameters $M^2$  are  negligible. The values of the Borel parameters $M^2$ are shown explicitly in Table 1.
From Figs.2-3 and Table 1, I can see that the Borel parameters $M^2$ in the QCD sum rules for the mass-shift $\delta m$ and decay-constant-shift $\delta f$ of the same meson are different. It is not un-acceptable, as the mass-shift $\delta m$ and decay-constant-shift $\delta f$ come from different QCD sum rules, not coupled QCD sum rules, see Eq.(39), the platforms maybe appear in different places  in different QCD sum rules.

I can obtain the shifts of the masses and decay constants of the heavy mesons in the nuclear matter in the Borel windows, which are shown explicitly in Table 2.
From the Table 2, I can obtain the fractions of the shifts $\frac{\delta m_{D/D^*/D_0/D_1}}{m_{D/D^*/D_0/D_1}}\leq 5\%$, $\frac{\delta f_{D/D^*/D_0/D_1}}{f_{D/D^*/D_0/D_1}}\leq 10\%$, $\frac{\delta m_{B/B^*/B_0/B_1}}{m_{B/B^*/B_0/B_1}}= (5-15)\%$ and $\frac{\delta f_{B/B^*/B_0/B_1}}{f_{B/B^*/B_0/B_1}}= (25-55)\%$, which are shown explicitly in Table 3.
In calculations, I observe that the main contributions come from the terms $m_c\langle\bar{q}q\rangle_N$ and   $m_b\langle\bar{q}q\rangle_N$.
 From Table 3, I can see that the next-to-leading order corrections $\alpha_s \langle\bar{q}q\rangle_N$ are important. In the case of  the shifts $\frac{\delta m_{B^*/B_1}}{m_{B^*/B_1}}$ and $\frac{\delta f_{B^*/B_1}}{f_{B^*/B_1}}$, the next-to-leading order contributions $\alpha_s \langle\bar{q}q\rangle_N$ and the leading order contributions $\langle\bar{q}q\rangle_N$ are almost equivalent. In this article, I choose the special energy scale $\mu=1\,\rm{GeV}$. The logarithm $\log \frac{m_b^2}{\mu^2}$ in the next-to-leading contributions is very large and enhances the next-to-leading  contributions greatly.  Although the nuclear matter induced condensates  evolve with the   renormalization group equation, their  evolving behaviors with the energy scales are not well known, as  this subject has not been studied in details yet at the present time. A larger energy scale $\mu$ can lead to smaller logarithm $\log \frac{m_b^2}{\mu^2}$ therefore  more reasonable predictions. In Table 4, I present the main uncertainties, which originate from the uncertainties of the heavy quark masses and the continuum threshold parameters.

\begin{table}
\begin{center}
\begin{tabular}{|c|c|c|c|c|c|c|c|c|c|}\hline\hline
       &$\delta m_{D}$ &$\delta m_{D^*}$ &$\delta m_{D_0}$  &$\delta m_{D_1}$  &$\delta m_{B}$  &$\delta m_{B^*}$  &$\delta m_{B_0}$  &$\delta m_{B_1}$\\ \hline
 $M^2$ &$4.4-5.4$      &$4.6-5.6$        &$6.0-7.0$         &$6.6-7.6$         &$29-33$         &$30-34$           &$32-36$           &$32-36$      \\ \hline\hline
       &$\delta f_{D}$ &$\delta f_{D^*}$ &$\delta f_{D_0}$  &$\delta f_{D_1}$  &$\delta f_{B}$  &$\delta f_{B^*}$  &$\delta f_{B_0}$  &$\delta f_{B_1}$\\ \hline
 $M^2$ &$1.9-2.9$      &$3.5-4.5$        &$4.3-5.3$         &$5.3-6.3$         &$25-29$         &$27-31$           &$30-34$           &$31-35$    \\ \hline\hline
\end{tabular}
\end{center}
\caption{ The Borel parameters in the QCD sum rules for the shifts of the masses and decay constants of the heavy mesons in the nuclear matter, the unit is $\rm{GeV}^2$.   }
\end{table}

\begin{table}
\begin{center}
\begin{tabular}{|c|c|c|c|c|c|c|c|c|c|}\hline\hline
           &$\delta m_{D}$ &$\delta m_{D^*}$ &$\delta m_{D_0}$  &$\delta m_{D_1}$  &$\delta m_{B}$  &$\delta m_{B^*}$  &$\delta m_{B_0}$  &$\delta m_{B_1}$\\ \hline
{\rm NLO}  &$-72$          &$-102$           &$80$              &$97$              &$-473$          &$-687$            &$295$             &$522$   \\ \hline
{\rm LO}   &$-47$          &$-70$            &$54$              &$66$              &$-329$          &$-340$            &$209$             &$260$   \\ \hline
\cite{Hay} &$-48$          &                 &                  &                  &                &                  &                  &\\ \hline
\cite{Hil} &$+45$          &                 &                  &                  & $+60$          &                  &                  &\\ \hline
\cite{Azi} &$-46$          &                 &                  &                  & $-242$         &                  &                  &\\ \hline\hline
           &$\delta f_{D}$ &$\delta f_{D^*}$ &$\delta f_{D_0}$  &$\delta f_{D_1}$  &$\delta f_{B}$  &$\delta f_{B^*}$  &$\delta f_{B_0}$  &$\delta f_{B_1}$\\ \hline
{\rm NLO}  &$-6$           &$-26$            &$11$              &$31$              &$-71$           &$-111$            &$56$              &$134$  \\ \hline
{\rm LO}   &$-4$           &$-18$            &$7$               &$21$              &$-48$           &$-55$             &$39$              &$67$  \\ \hline
\cite{Azi} &$-2$           &                 &                  &                  & $-23$          &                  &                  &\\ \hline\hline
\end{tabular}
\end{center}
\caption{ The shifts of the masses and decay constants of the heavy mesons in the nuclear matter, where the NLO (LO) denotes contributions up to  the next-to-leading order (leading order) are included, the unit is MeV.  }
\end{table}

\begin{table}
\begin{center}
\begin{tabular}{|c|c|c|c|c|c|c|c|c|c|}\hline\hline
           &$\frac{\delta m_{D}}{m_{D}}$   &$\frac{\delta m_{D^*}}{m_{D^*}}$  &$\frac{\delta m_{D_0}}{m_{D_0}}$   &$\frac{\delta m_{D_1}}{m_{D_1}}$
           &$\frac{\delta m_{B}}{ m_{B}}$  &$\frac{\delta m_{B^*}}{m_{B^*}}$  &$\frac{\delta m_{B_0}}{m_{B_0}}$   &$\frac{\delta m_{B_1}}{m_{B_1}}$\\ \hline
{\rm NLO}  &$-4\%$         &$-5\%$           &$3\%$             &$4\%$             &$-9\%$          &$-13\%$           &$5\%$             &$9\%$   \\ \hline
{\rm LO}   &$-3\%$         &$-3\%$           &$2\%$             &$3\%$             &$-6\%$          &$-6\%$            &$4\%$             &$5\%$   \\  \hline\hline
           &$\frac{\delta f_{D}}{f_{D}}$   &$\frac{\delta f_{D^*}}{f_{D^*}}$  &$\frac{\delta f_{D_0}}{f_{D_0}}$   &$\frac{\delta f_{D_1}}{f_{D_1}}$
           &$\frac{\delta f_{B}}{f_{B}}$   &$\frac{\delta f_{B^*}}{f_{B^*}}$  &$\frac{\delta f_{B_0}}{f_{B_0}}$   &$\frac{\delta f_{B_1}}{f_{B_1}}$\\ \hline
{\rm NLO}  &$-3\%$         &$-10\%$          &$6\%$             &$10\%$            &$-37\%$         &$-57\%$           &$24\%$            &$53\%$  \\ \hline
{\rm LO}   &$-2\%$         &$-7\%$           &$4\%$             &$7\%$             &$-25\%$         &$-28\%$           &$17\%$            &$26\%$  \\ \hline\hline
\end{tabular}
\end{center}
\caption{ The fractions of the  shifts of the masses and decay constants of the heavy mesons in the nuclear matter, where the NLO (LO) denotes contributions up to  the next-to-leading order (leading order) are included.  }
\end{table}

\begin{table}
\begin{center}
\begin{tabular}{|c|c|c|c|c|c|c|c|c|c|}\hline\hline
             &$\delta (\delta m_{D})$   &$\delta(\delta m_{D^*})$   &$\delta(\delta m_{D_0})$     &$\delta(\delta m_{D_1})$
             &$\delta(\delta m_{B})$    &$\delta(\delta m_{B^*})$   &$\delta(\delta m_{B_0})$     &$\delta(\delta m_{B_1})$              \\ \hline
$\delta m_Q$ &$\pm14$                   &$\pm4$                     &$\pm26$                      &$\pm6$
             &$\pm18$                   &$\pm1$                     &$\pm25$                      &$\pm1$   \\ \hline
$\delta s_0$ &$\pm9$                    &$\pm14$                    &$\pm12$                      &$\pm13$
             &$\pm65$                   &$\pm80$                    &$\pm39$                      &$\pm70$   \\ \hline \hline
             &$\delta (\delta f_{D}$)   &$\delta (\delta f_{D^*})$  &$\delta (\delta f_{D_0})$    &$\delta (\delta f_{D_1})$
             &$\delta (\delta f_{B})$   &$\delta (\delta f_{B^*})$  &$\delta (\delta f_{B_0})$    &$\delta (\delta f_{B_1})$\\ \hline
$\delta m_Q$ &$\pm1$                    &$\pm1$                     &$\pm3$                       &$\pm2$
             &$\pm2$                    &$\pm1$                     &$\pm3$                       &$\pm0$   \\ \hline
$\delta s_0$ &$\pm1$                    &$\pm7$                     &$\pm4$                       &$\pm8$
             &$\pm21$                   &$\pm25$                    &$\pm15$                      &$\pm34$   \\ \hline \hline
\end{tabular}
\end{center}
\caption{ The uncertainties of the shifts of
    the masses and decay constants of the heavy mesons in the nuclear matter originate from the uncertainties of the heavy quark masses and continuum threshold parameters, where the unit is MeV.  }
\end{table}

The mass-shifts of the negative (positive) parity  mesons are negative (positive), the decays of the high
charmonium  states to the $D\bar{D}$ and $D^*\bar{D}^*$ ($D_0\bar{D}_0$ and $D_1\bar{D}_1$) pairs are  enhanced (suppressed) in the phase space,
 and we  should take into account those effects carefully in studying the production of the  $J/\psi$ so as to identifying the quark-gluon plasmas.

 The  currents $\bar{Q}  q$ and $\bar{Q}i\gamma_5 q$ (also $\bar{Q}\gamma_\mu q$ and $\bar{Q}\gamma_\mu\gamma_5 q$) are mixed with each other under the  chiral transformation $q\to e^{i\alpha\gamma_5}q$, the  currents $\bar{Q}  q$, $\bar{Q}i\gamma_5 q$, $\bar{Q}\gamma_\mu q$, $\bar{Q}\gamma_\mu\gamma_5 q$ are not conserved in the limit $m_q \to 0$, it is better to take the doublets $(D,D_0)$ and $(D^*,D_1)$ as  the parity-doublets rather than  the chiral-doublets.
The  quark condensate $\langle\bar{q}q\rangle_{\rho_N}$ serves as the order parameter,
  and  undergoes  reduction in the nuclear matter,
    the chiral symmetry is partially  restored; however, there appear new medium-induced condensates, which also break the chiral symmetry. In this article, the $\langle
{\cal{O}}\rangle_N$  are companied by  the heavy quark masses $m_Q$, $m_Q^2$ or $m_Q^3$, the net effects cannot warrant that  the chiral symmetry is monotonously restored with the increase of the $\rho_N$.
 When the $\rho_N$ is large enough, the order parameter $\langle\bar{q}q\rangle_{\rho_N} \to 0$, the chiral symmetry is restored, the Fermi  gas  approximation for the nuclear matter breaks down,  and the parity-doublets
  maybe have degenerated  masses approximately. In this article, I study the parity-doublets  at the low $\rho_N$, the mass breaking effects of the parity-doublets   maybe even larger, see Table 2.
  We expect that smaller mass splitting of the parity-doublets at the high nuclear density is favored, however, larger  mass splitting of the parity-doublets at the lower nuclear density cannot be excluded.
  In Refs.\cite{Hil,Hilger2}, the mass center $\overline{m}_P$ of the pseudoscalar mesons increases in the nuclear matter while the mass center $\overline{m}_S$ of the scalar mesons  decreases in the nuclear matter,  the mass breaking effect  $\overline{m}_S-\overline{m}_P$ of the parity-doublets is  smaller than that in the vacuum.

In Table 5, I show the scattering lengths $a_{D}$,  $a_{D^*}$,  $a_{D_0}$,  $a_{D_1}$,  $a_{B}$,  $a_{B^*}$,   $a_{B_0}$,  $a_{B_1}$ explicitly, the  $a_{D}$,  $a_{D^*}$,    $a_{B}$,  $a_{B^*}$ are negative, which indicate the interactions $DN$,  $D^*N$,    $BN$,  $B^*N$ are attractive,  the  $a_{D_0}$,  $a_{D_1}$,     $a_{B_0}$,  $a_{B_1}$ are positive, which indicate the interactions $D_0N$,  $D_1N$,    $B_0N$,  $B_0N$ are repulsive. It is difficult (possible)  to form the $D_0N$,  $D_1N$,    $B_0N$,  $B_0N$ ($DN$,  $D^*N$,    $BN$,  $B^*N$) bound states. In Ref.\cite{ZhangDN}, Zhang studies the $S$-wave $DN$ and $D^*N$ bound states with the QCD sum rules, the numerical results indicate that  the $\Sigma_c(2800)$  and $\Lambda_c(2940)$   can  be assigned to be the $S$-wave $DN$  state with $J^P ={\frac{1}{2}}^-$ and  the $S$-wave $D^*N$  state with $J^P ={\frac{3}{2}}^-$,  respectively.

\begin{table}
\begin{center}
\begin{tabular}{|c|c|c|c|c|c|c|c|c|c|}\hline\hline
           &$a_{D}$        &$a_{D^*}$        &$a_{D_0}$         &$a_{D_1}$         &$a_{B}$         &$a_{B^*}$         &$a_{B_0}$         &$a_{B_1}$\\ \hline
{\rm NLO}  &$-1.1$         &$-1.5$           &$1.3$             &$1.6$             &$-8.9$          &$-12.9$           &$5.6$             &$9.9$   \\ \hline
{\rm LO}   &$-0.7$         &$-1.1$           &$0.9$             &$1.1$             &$-6.2$          &$-6.4$            &$4.0$             &$5.0$   \\ \hline\hline
\end{tabular}
\end{center}
\caption{ The heavy-meson-nucleon scattering lengths, where the NLO (LO) denotes contributions up to  the next-to-leading order (the leading order) are included, the unit is fm.  }
\end{table}

\begin{figure}
 \centering
 \includegraphics[totalheight=6cm,width=8cm]{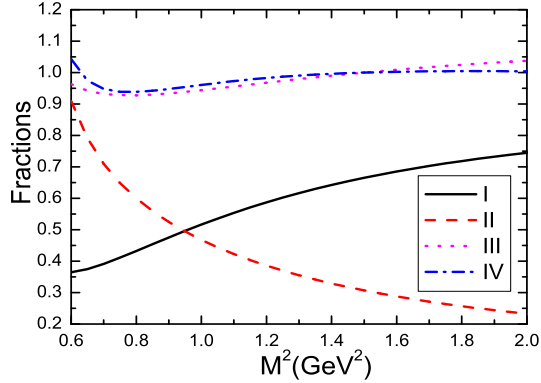}
   \caption{(Color online) The contributions of the perturbative term (I) and quark condensate term (II) in the QCD sum rules for the $D$ mesons  in the vacuum. Furthermore, I  show  the  mass $m_D$ (III) and decay constant $f_D$ (IV) explicitly, which are normalized to be 1 at the value $M^2=1.5\rm{GeV}^2$. }
\end{figure}

In the present work and Refs.\cite{Hay,WangHuang,Azi},  the correlation functions  are  divided
into  a vacuum part   and a static one-nucleon part, and  the nuclear matter  induced effects are extracted  explicitly;
  while in Refs.\cite{Hil,Hilger2},   the pole terms of (or ground state contributions to)  the  hadronic spectral densities of the whole correlation functions are  parameterized as $\Delta\Pi(\omega)= F_{+}\delta(\omega-m_{+})- F_{-}\delta(\omega+m_{-})$, where $m_{\pm}=m\pm\Delta m$ and $F_{\pm}=F\pm\Delta F$,
and  QCD  sum rules for the   mass center $\overline{m}$ and the mass splitting $\Delta m$ are obtained.
In the leading order approximation, the present predictions of  the $\delta m_{D}$ and $\delta m_{B}$ are compatible with that of Refs.\cite{Hay,Azi} and differ greatly from that of Refs.\cite{Hil,Hilger2}, see Table 2.
The values obtained from the QCD sum rules depend heavily on the Borel windows, the values extracted from different Borel windows especially in different QCD sum rules maybe differ from each other greatly.

In Refs.\cite{Hil,Hilger2}, the authors study the masses of the heavy mesons in the nuclear matter directly by  including both the vacuum part   and the static one-nucleon part in the QCD sum rules, then the continuum contributions are well approximated by $\rho_{QCD}(\omega^2)\theta\left(\omega^2-{\omega_0^{\pm}}^2\right)$, where the ${\omega_0^{\pm}}^2$ are the continuum threshold parameters,   it is one of the advantages of Refs.\cite{Hil,Hilger2}. However, they define the moments $S_n(M^2)$ to study the mass-shifts,
\begin{eqnarray}
S_n(M^2)&=&\int_{\omega_0^-}^{\omega_0^+}d\omega \omega^n \Delta\Pi(\omega) \exp\left( -\frac{\omega^2}{M^2}\right)\, ,
\end{eqnarray}
the odd moment $o=S_0(M^2)$ and the even moment $e=S_1(M^2)$, then obtain $\frac{do}{d(1/M^2)}=-S_2(M^2)$ and $\frac{de}{d(1/M^2)}=-S_3(M^2)$ by assuming the $F_{\pm}$ and $m_{\pm}$ are independent on the Borel parameters  at the phenomenological side.  In fact, $\frac{do}{d(1/M^2)}\neq-S_2(M^2)$ and $\frac{de}{d(1/M^2)}\neq-S_3(M^2)$ at the operator product expansion side according to the QCD spectral densities $\Delta\Pi(\omega)$, which  depend  on the Borel parameters explicitly, the approximations $\frac{do}{d(1/M^2)}=-S_2(M^2)$ and $\frac{de}{d(1/M^2)}=-S_3(M^2)$ lead to undetermined uncertainties.

In Refs.\cite{Hil,Hilger2}, the perturbative $\mathcal{O}(\alpha_s)$ corrections  to the  perturbative terms are taken into account. In the QCD sum rules for the pseudoscalar $D$  mesons  in the vacuum, if we take into account the perturbative  $\mathcal{O}(\alpha_s)$ corrections to the perturbative term and vacuum condensate term, the  two criteria (pole dominance and convergence of the operator product expansion) of the QCD sum rules leads  to the Borel window $M^2=(1.2-1.8)\,\rm{GeV}^2$, the resulting predictions
of the mass $m_{D}$  and decay constant $f_{D}$  are consistent with the experimental data.  In Fig.4, I plot the contributions of the perturbative term   and quark condensate  term in the operator product expansion. From the figure, I can see that the main contributions come from the perturbative term, the quark condensate $\langle \bar{q}q\rangle$ plays a less important role.

The modifications of the condensates in the nuclear matter are mild, for example, $\langle\bar{q}q\rangle_{\rho_N}\approx 0.64 \langle\bar{q}q\rangle $, while the perturbative contributions are not modified (or modified slightly by introducing a minor  splitting $\Delta s_0$, ${\omega_0^{\pm}}^2=s_0\pm\Delta s_0$) by the nuclear matter.
If we turn on the in-medium effects, the contributions of  the quark condensate  are even smaller, the Borel windows are determined dominantly
by the perturbative terms \cite{Hil,Hilger2}.
If the     perturbative  $\mathcal{O}(\alpha_s^2)$ corrections to the perturbative terms are also included, the contributions of the perturbative are even larger \cite{WangJHEP}, the QCD sum rules are  dominated by the perturbative terms, which are not (or slightly) affected by the nuclear matter.
It is not favored to extract the  mass-shifts in the nuclear matter, and impairs the predictive ability.

In the present work and Refs.\cite{Hay,WangHuang,Azi},  the correlation functions   are divided into
   the vacuum part   and the static one-nucleon part,  which are of the orders ${\mathcal{O}}(0)$ and ${\mathcal{O}}(\rho_N)$, respectively. We can obtain  independent QCD sum rules from the two parts respectively. The QCD sum rules correspond to the orders ${\mathcal{O}}(0)$ and ${\mathcal{O}}(\rho_N)$ respectively can   have quite different Borel parameters. In this article, I  separate the nuclear matter induced effects unambiguously, study the QCD sum rules correspond to the order ${\mathcal{O}}(\rho_N)$, and determine the Borel parameters by the criteria of the QCD sum rules.

In the conventional QCD sum rules, we usually choose the  Borel parameters $M^2$  to satisfy the  following three  criteria:

$\bf{1_\cdot}$ Pole dominance at the phenomenological side;

$\bf{2_\cdot}$ Convergence of the operator product expansion;

$\bf{3_\cdot}$ Appearance of the Borel platforms.

In the present work and Refs.\cite{Hay,WangHuang,Azi}, the nuclear matter  induced effects are extracted  explicitly, the resulting QCD sum rules are not contaminated by the contributions of the vacuum part, the Borel windows are determined completely by the nuclear matter  induced effects,  it  is the advantage.
As the QCD spectral densities are of the form $\delta(\omega^2-m_Q^2)$, we have to take the hadronic spectral densities to be the  form $\delta(\omega^2-m_{H}^2)$ and model the continuum contributions with the function $\delta(\omega^2-s_0)$, and determine the $s_0$ by some constraints, see Eq.(16), where the $H$ denotes  the ground state and excited state heavy mesons. In this article, I attribute the higher excited states to the continuum contributions, the $\delta$-type hadronic spectral densities  make sense. So the pole dominance at the phenomenological side can be  released as the  continuum contributions are already taken into account.
Furthermore, I expect that the couplings of
 the interpolating  currents  to the excited states are more weak than that to the ground states, the uncertainties originate from continuum contributions are very small. For example, the decay constants of the pseudoscalar mesons $\pi(140)$ and $\pi(1300)$ have the hierarchy $f_{\pi(1300)}\ll f_{\pi(140)}$ from the Dyson-Schwinger equation \cite{CDRoberts}, the lattice QCD \cite{Latt-pion},  the QCD sum rules \cite{QCDSR-pion}, etc, or from the experimental data \cite{pion-exp}.

In the present work and Refs.\cite{Hay,WangHuang,Azi}, large Borel parameters are chosen  to warrant the convergence of the operator product expansion and to obtain the Borel platforms, and small Borel parameters cannot lead to platforms. In the Borel windows, where the platforms appear, the main contributions come from terms $\langle \bar{q}q\rangle_N$, the operator product expansion is well convergent. The criteria $\bf{2}$ and $\bf{3}$ can be satisfied.  The continuum contributions are not  suppressed efficiently for large Borel parameters compared to that for small Borel parameters.   In calculations, I  observe that the predictions are insensitive to the $s_0$, the uncertainties originate from the continuum threshold parameters $s_0$ are very small in almost all cases,  the large Borel parameters make sense. Furthermore, the    continuum contributions are already taken into account. On the other hand, from Eqs.(8-9) and Eqs.(12-13), we can see that the mass-shifts $\delta m_{D/D_0/D^*/D_1}$ and decay constant shifts $\delta f_{D/D_0/D^*/D_1}$ reduce to zero in the limit $\rho_N \to 0$, the QCD sum rules correspond to the nuclear matter induced effects decouple, their Borel parameters (irrespective of large or small) are also irrelevant to the ones in the QCD sum rules for the vacuum part of the correlation functions.
So the present predictions are sensible.

The predictions depend on the in-medium hadronic spectral functions \cite{Kwon-2008}, for example, there are two generic prototypes of the  in-medium
spectral functions for the $\rho$ meson, they differ in details at the
low mass end of the spectrum. The Klingl-Kaise-Weise spectral function
 emphasizes the role of chiral in-medium $\pi\pi$ interactions \cite{KKW-97}, while  the Rapp-Wambach spectral function focuses
on the role of nucleon-hole, $\Delta(1232)$-hole and $N^*(1520)$-hole excitations \cite{RW-99}. Both of the
spectral functions account quite well for the low-mass enhancements observed in dilepton spectra from high-energy nuclear collisions. However, the QCD
sum rules analysis of the lowest spectral moments reveals qualitative differences with respect to their Brown-Rho  scaling properties \cite{Kwon-2008}. If the simple spectral densities $F\delta(\omega^2-M_{P/V}^2)$ analogous to the ones in Refs.\cite{Hil,Hilger2} are taken, where the $P$ denotes the pseudoscalar mesons $\pi$, $\eta_c$, the $V$ denotes the vector mesons $\rho$, $\omega$, $\phi$, $J/\psi$, the $F$ denotes the constant pole residues, the in-medium  mass-shifts $\delta M_{P/V}$ are smaller  than zero qualitatively \cite{Lee-Vector}. I expect that the S-wave mesons $q^{\prime}\bar{q}$, $c\bar{q}$, $c\bar{c}$ with the spin-parity $J^P=0^-$ (or $1^-$) have analogous in-medium mass-shifts, at least qualitatively. Further studies based on more sophisticated hadronic spectral densities are needed.

In fact, there are controversies about the mass-shifts of the $D$ and $B$ mesons in the nuclear matter, some  theoretical approaches indicate negative mass-shifts \cite{DN-BSE-Negative}, while others indicate positive mass-shifts \cite{Positive-BD}.  The different predictions  originate mainly from whether or not the   heavy pseudoscalar and heavy vector mesons are treated  on equal footing in the coupled-channel approaches. If we obtain   the meson-baryon interaction kernel  by   treating the heavy pseudoscalar and heavy vector mesons on equal footing as required by heavy quark symmetry, the  mass-shift $\delta M_{D}$ is negative \cite{DN-BSE-Negative}, which is consistent with the present work;
furthermore, the attractive D-nucleus interaction can lead to the formation  of $D$-nucleus bound states, which can be confronted to the experimental data in the future directly \cite{D-Nuclei}.

The upcoming  FAIR  project at GSI
 provides the opportunity to study the in-medium properties of the charmoniums  or charmed hadrons for the first time, however,
the high mass of charmed hadrons requires a high momentum in the antiproton beam to produce them, the
conditions for observing in-medium effects seem unfavorable, as  the hadrons  sensitive
to the in-medium effects are either at rest or have a small momentum relative to the nuclear
medium. We have to find
processes that would slow down the charmed hadrons inside the nuclear matter, but this
requires more detailed theoretical studies.
Further theoretical studies on the reaction
dynamics and on the exploration of the experimental
ability  to identify more complicated
processes are still needed.

\section{Conclusion}
 In this article, I divide  the two-point correlation functions of the  scalar, pseudoscalar, vector and axialvector currents in the nuclear matter
into  two parts, i.e. the vacuum part   and the static one-nucleon part, then study  the in-medium modifications of  the masses and decay constants by deriving
 QCD sum rules from the   static one-nucleon part of the two-point correlation functions. In the operator product expansion,
 I calculate the contributions of the nuclear matter induced condensates up to dimension 5,  especially I calculate the next-to-leading order contributions of the in-medium quark condensate and obtain concise expressions, which also have applications in studying the mesons properties in the vacuum.
  In calculation, I observe that the next-to-leading order contributions of the in-medium quark condensate are very large and should be taken into account.

 All in all, I study the properties of the scalar, pseudoscalar, vector and axialvector heavy mesons  with the QCD sum rules in a systematic way,  and obtain    the shifts of the masses and decay constants in the nuclear matter.
 The numerical results indicate that the mass-shifts of the negative parity and positive parity heavy mesons are negative and positive, respectively. For the pseudoscalar   meson $D$, I obtain the prediction $\delta M_{D}<0$, which is in contrast to the prediction in Refs.\cite{Hil,Hilger2}, where the mass-shift is positive  $\delta M_{D}>0$.   In Refs.\cite{Hil,Hilger2}, the authors study the masses of the heavy mesons in the nuclear matter directly by  including both the vacuum part   and
  static one-nucleon part with  the QCD sum rules, and   parameterize the spectral density of the whole correlation functions by a simple function
  $\Delta\Pi(\omega)= F_{+}\delta(\omega-m_{+})- F_{-}\delta(\omega+m_{-})$.
  I discuss the differences between the QCD sum rules in the present work and that in Refs.\cite{Hil,Hilger2} in details, and show why I prefer the present predictions.   In the present work and Refs.\cite{Hay,WangHuang,Azi,Hil,Hilger2}, the finite widths of the mesons in the nuclear matter are neglected, further   studies based on the  more sophisticated hadronic spectral densities by including the finite widths are  needed.

  As the masses of the heavy meson paries, such as the $D\bar{D}$, $D^*\bar{D}^*$, $D_0 \bar{D}_0$, $D_1 \bar{D}_1$ are modified in the nuclear environment,
  we  should take into account those effects carefully in studying the production of the  $J/\psi$ (and $\Upsilon$) so as to identifying the quark-gluon plasmas. Furthermore, I study the heavy-meson-nucleon scattering lengths as a byproduct,
and obtain the conclusion qualitatively about the possible existence  of the  heavy-meson-nucleon bound states.

\section*{Acknowledgements}
This  work is supported by National Natural Science Foundation,
Grant Numbers 11375063, and Natural Science Foundation of Hebei province, Grant Number A2014502017.

\end{document}